\documentclass[trackchanges,twocolumn]{aastex701}

\newcommand{\mwest}{\textsc{MWest}}
\newcommand{\symphony}{\textsc{Symphony MilkyWay}}
\newcommand{\expcode}{\textsc{EXP }}
\usepackage{comment}
\let\tablenum\relax
\newcommand{\snum}[1]{#1\relax} % wrapper that appends \relax automatically
\usepackage{siunitx} % in preamble
\usepackage[x11names]{xcolor}
 
\usepackage{amsmath}
\shorttitle{\textsc{The response to the LMC's passage in cosmological context}}
\shortauthors{Darragh-Ford, Garavito-Camargo, et. al.,}

\begin{document}

\title{Shaping the Milky Way. II. The dark matter halo's response to the LMC's passage in a cosmological context}

\author[orcid=0000-0002-8800-5652]{Elise Darragh-Ford}
\email{edarragh@stanford.edu}
\affiliation{Kavli Institute for Particle Astrophysics \& Cosmology, P. O. Box 2450, Stanford University, Stanford, CA 94305, USA}
\affiliation{Department of Physics, Stanford University, 382 Via Pueblo Mall, Stanford, CA 94305, USA}
\affiliation{SLAC National Accelerator Laboratory, Menlo Park, CA 94025, USA}

\author[orcid=0000-0001-7107-1744]{Nicolás Garavito-Camargo}\thanks{NASA NFHP Einstein Fellow}
\affiliation{Steward Observatory, University of Arizona, 933 North Cherry Avenue, Tucson, AZ 85721, USA}
\affiliation{Department of Astronomy, University of Maryland, College Park, MD 20742, USA}
\email[show]{garavito@umd.edu}
\correspondingauthor{Nicolás Garavito-Camargo}

\author[orcid=0000-0002-8354-7356]{Arpit Arora}
\email{arora125@sas.upenn.edu}
\affiliation{Department of Astronomy, University of Washington, Seattle, WA 98195, USA}

\author[orcid=0000-0003-2229-011X]{Risa H. Wechsler}
\affiliation{Kavli Institute for Particle Astrophysics \& Cosmology, P. O. Box 2450, Stanford University, Stanford, CA 94305, USA}
\affiliation{Department of Physics, Stanford University, 382 Via Pueblo Mall, Stanford, CA 94305, USA}
\affiliation{SLAC National Accelerator Laboratory, Menlo Park, CA 94025, USA}
\email{rwechsler@stanford.edu}

\author[orcid=0000-0001-9863-5394]{Phil Mansfield}
\affiliation{Kavli Institute for Particle Astrophysics \& Cosmology, P. O. Box 2450, Stanford University, Stanford, CA 94305, USA}
\affiliation{SLAC National Accelerator Laboratory, Menlo Park, CA 94025, USA}
\email{phil1@stanford.edu}

\author[orcid=0000-0003-0715-2173]{Gurtina Besla}
\affiliation{Steward Observatory, University of Arizona, 933 North Cherry Avenue, Tucson, AZ 85721, USA}
\email{gbesla@arizona.edu}

\author[0000-0003-1517-3935]{Michael S. Petersen}
\affiliation{Institute for Astronomy, University of Edinburgh, Royal Observatory, Blackford Hill, Edinburgh EH9 3HJ, UK}
\email{michael.petersen@roe.ac.uk}

\author[0000-0003-2660-2889]{Martin D. Weinberg}
\affiliation{Department of Astronomy, University of Massachusetts, Amherst, MA 01003-9305}
\email{weinberg@astro.umass.edu}

\author[orcid=0000-0003-3213-8736]{Silvio Varela-Lavin}
\affiliation{Departamento de F\'isica, Universidad Tecnica Federico Santa Maria, Avenida Espa\~na 1680, Valpara\'iso, Chile}
\email{silvio.varela@userena.cl}

\author[orcid=0000-0002-6635-4712]{Deveshi Buch}
\affiliation{Kavli Institute for Particle Astrophysics \& Cosmology, P. O. Box 2450, Stanford University, Stanford, CA 94305, USA}
\affiliation{Department of Computer Science, Stanford University, 353 Jane Stanford Way, Stanford, CA 94305, USA}
\email{deveshi@cs.stanford.edu}

\author[orcid=0000-0002-6993-0826]{Emily C. Cunningham}
\affiliation{Department of Astronomy, Boston University, 725 Commonwealth Avenue, Boston, MA 02215, USA}
\email{eccunnin@bu.edu}

\author[orcid=0000-0003-2594-8052]{Kathryne J. Daniel}
\affiliation{Steward Observatory, University of Arizona, 933 North Cherry Avenue, Tucson, AZ 85721, USA}
\email{kjdaniel@arizona.edu}

\author[orcid=0000-0002-6810-1110]{Facundo A. Gómez}
\affiliation{Departamento de Astronom\'ia, Universidad de La Serena, Av. Raúl Bitrán 1305, La Serena, Chile}
\email{fagomez@userena.cl}

\author[orcid=0000-0001-6244-6727]{Kathryn V. Johnston}
\affiliation{Department of Astronomy, Columbia University, 550 West 120th Street, New York, NY 10027, USA}
\email{kvj@astro.columbia.edu}

\author[orcid=0000-0002-6810-1110]{Chervin F. P. Laporte}
\affiliation{LIRA, Observatoire de Paris, Universit\'e PSL, Sorbonne Universit\'e, Universit\'e Paris Cit\'e, CY Cergy Paris Universit\'e, CNRS, 92190 Meudon, France}
\affiliation{Institut de Ciencies del Cosmos (ICCUB), Universitat de Barcelona (IEEC-UB), Martí i Franquès 1, E-08028 Barcelona, Spain}
\affiliation{Kavli IPMU (WPI), UTIAS, The University of Tokyo, Kashiwa, Chiba 277-8583, Japan}
\email{chervin.laporte@icc.ub.edu}

\author[orcid=0000-0002-1200-0820]{Yao-Yuan Mao}
\affiliation{Department of Physics and Astronomy, University of Utah, Salt Lake City, UT 84112, USA}
\email{yymao@astro.utah.edu}

\author[orcid=0000-0002-1182-3825]{Ethan O.~Nadler}
\affiliation{Department of Astronomy \& Astrophysics, University of California, San Diego, La Jolla, CA 92093, USA}
\email{enadler@ucsd.edu}

\author[orcid=0000-0003-3939-3297]{Robyn Sanderson}
\affiliation{Department of Physics \& Astronomy, University of Pennsylvania, 209 S 33rd St, Philadelphia, PA 19104, USA}
\email{robynes@sas.upenn.edu}

\collaboration{all}{The EXP collaboration}
\noaffiliation 


%% the nearby authors are not part of the surrounding collaborations.

\begin{abstract}
The distribution of dark matter in the Milky Way (MW) is expected to exhibit a large-scale dynamical response to the recent infall of the LMC. This event produces a dynamical friction wake and shifts the MW's halo density center. The structure of this response encodes information about the LMC-MW mass ratio, the LMC's orbit, the MW halo's pre-infall structure and could provide constraints on dark matter physics. To extract this information, a method to separate these effects and recover the initial shape of the MW's halo is required. Here, we use basis function expansions to analyze the halo response in eighteen simulations of MW–LMC–like interactions from the \mwest\ cosmological, dark-matter-only zoom-in simulations. The results show that mergers similar to the LMC consistently generate a significant dipole and a secondary quadrupole response in the halo. The dipole arises from the host density center displacement and halo distortions, and its amplitude scales as the square of the MW--LMC mass ratio, peaking 0.2-0.7 Gyr after the LMC's pericenter. The quadrupole's strength depends primarily on the original axis ratios of the host halo, though contributions from the dynamical friction wake cause it to peak less than 0.3 Gyr before pericenter. Future measurements of both the dipole and quadrupole imprints of the LMC's passage in the density of the MW's stellar halo should be able to disentangle these effects and provide insight into the initial structure of the MW's halo, the MW's response, and the mass of the LMC.
\end{abstract}

%\keywords{}


%% and \citet commands to identify citations. The citations are

\section{Introduction}\label{sec:intro}

The infall of massive satellites leaves long-lived dynamical signatures in their host halos, offering a direct probe of the satellite's mass and orbit and the host's structure. Among Local Group satellites, the Large Magellanic Cloud (LMC) is unique: with a mass of $\sim$10\% of the Milky Way (MW), and likely on its first infall \citep{besla07, Kallivayalil13, Watkins_24}, it is expected to induce large-scale perturbations in the MW halo --- the \textit{LMC wake} \citep[e.g.,][]{Weinberg98, Laporte18, garavito_19, Petersen_20, GC21, Lilleengen_23}. These distortions in the dark matter may also be visible in the stellar halo, and provide a new avenue to understand the dynamics of the MW--LMC interaction, measure the MW--LMC mass ratio, and test the response of dark matter (DM) halos to strong perturbations.

The halo response can be separated into two main effects. The \textit{dynamical friction wake}, first described by \citet{Chandrasekhar1943}, trails the orbit of the satellite and depends sensitively on the properties of the DM particle and model \citep[e.g., ][]{lancaster20, Foote_23, Glennon_24}. The second is a displacement of the host's center of mass (COM), or ``collective response,'' in which the inner halo is displaced relative to the outer halo, producing a large-scale dipole \citep{Weinberg1999, Gomez15, Ogiya_16, GC21, Weinberg_23}. Both effects persist for several dynamical times, with maximum amplitude near the satellite's first pericentric passage.

Evidence for the LMC's wake has begun to emerge. Recent maps of the MW stellar halo show overdensities trailing the LMC orbit consistent with the dynamical friction wake \citep{belokurov19, Conroy_2021, Fushimi_23, Amarante_24, Cavieres25}. The collective response is more difficult to distinguish: outer halo overdensities may also arise from stellar substructure \citep{Cunninghnam_20, Amarante_24}, but velocity signatures of a displaced halo have recently been detected \citep{Petersen_21, Erkal_21, Yaaqib_24, Bystrom24, Chandra24, Yaaqib25}. The impact of the LMC is also seen in stellar streams and in the asymmetric distribution of ultra-faint satellites \citep{Shipp19, Koposov19, Nadler_2020, Arora24}. For a comprehensive review, see \citet{Vasiliev_23review}.

Most theoretical work on the wake to date has relied on idealized $N$-body simulations 
\citep[e.g.,][]{Mastropietro05, Gomez15, Laporte18, Petersen_20, GC21, 
Vasiliev_21, Lilleengen_23, vasiliev2023dear, Sheng_24}. 
For example, in idealized MW--LMC simulations, the dynamical friction wake induces an overdensity of 
DM particles of $\approx$30\% relative to an unperturbed MW halo. However, the orientation and amplitude of the COM displacement and the dynamical friction wake 
are sensitive to the MW--LMC mass ratio, the orbit of the LMC, and the MW's halo shape prior to the LMC's infall
\citep{Sheng_24}. However, cosmological halos are more complex; they exhibit ongoing accretion, substructure, and triaxial shapes \citep[e.g.,][]{Allgood06, Valluri_21, Ash23}, raising the question of how robust these predictions are in realistic environments.

In this paper, the second in a series using basis function expansions
(BFE) to quantify halo responses \citep{Arora25}, we investigate the MW 
halo's response to the LMC in a cosmological context. Using the \mwest\ 
suite of zoom-in cosmological simulations of MW--LMC analogs, we ask: 
(i) Are the dynamical friction wake and COM displacement predicted by 
idealized models also present in realistic cosmological halos? 
(ii) How can the contributions of these two effects be disentangled in 
the present-day halo response? 
(iii) What do these signatures reveal about the MW--LMC mass ratio and 
the underlying halo shape? 

We use BFE as our primary tool, as it provides a natural framework to
decompose halo responses into harmonics and connect them to physical 
mechanisms. A detailed discussion of BFE is presented in 
Section~\ref{sec:bfe}. Section~\ref{sec:methods} also describes the 
\mwest\ and \symphony\ simulations and our methods. 
Section~\ref{sec:results} presents the main results, including the 
harmonic decomposition of the LMC wake and COM displacement. We discuss 
their implications in Section~\ref{sec:discussion} and conclude in 
Section~\ref{sec:conclusions}.

\section{Simulations and Methods}\label{sec:methods}

Cosmological hydrodynamical simulations have been very successful at reproducing many of
the present-day observed properties of galaxies \citep{Volgesberger20}, 
and are becoming increasingly realistic in
capturing the physical processes that drive the dynamical evolution of a galaxy. However, a general framework that connects the simulated galaxy evolution to the dynamical theory of self-gravitating systems is still lacking.

BFE provides that connection --- a 
language to describe the galactic dynamics. By projecting the simulation particle phase-space
coordinates onto a set of basis 
functions, the self-gravitating system is decomposed into its
dynamically coherent structure, as in analytical perturbation theory \citep{Weinberg1999}. In the BFE framework, one follows the evolution of the leading harmonic modes (e.g., the monopole and dipole terms) to characterize the system's dynamical state. This framework can then easily facilitate comparisons across many halos and across different DM models. BFE can be used to simulate the galaxy's evolution and to analyze preexisting $N$-body simulations. Here we use BFE to decompose the response of a statistical sample of CDM, MW-mass halos to LMC-like satellites in the \mwest\ and \symphony\ suites. 

In the following subsections, we describe the simulations and methods used. We summarize the properties of the \mwest\ and \symphony\ simulations in Section~\ref{sec:simulations}. We describe how the particles of each halo are assigned by the halo finder in Section~\ref{sec:halos}. In Section~\ref{sec:orbits}, we give the properties of the LMC analogs and the choice of reference frames. In Section~\ref{sec:bfe}, we give an overview of BFE and describe how we use it in cosmological simulations, and in Section~\ref{sec:inertia_tensor}, we describe the shape-measurement method.

\subsection{The \mwest\ and \symphony\ suites}\label{sec:sims}

\subsubsection{Simulations}\label{sec:simulations}
The Milky Way-est (\mwest) suite is a set of twenty\footnote{Two halos from~\cite{buch2024milky} (\textsc{Halo 453} and \textsc{Halo 476}) have been removed due to particle data integrity issues. We do not expect 
them to significantly alter this analysis.} dark-matter-only zoom-in simulations of MW-like halos.
They are selected to have
$1 \times 10^{12}\, M_\odot < M_{\rm MW} < 1.8 \times 10^{12}\, M_\odot$ 
and $7 < c_{\rm host} < 16$, and MW-like accretion histories. 
Each host includes a Gaia--Sausage--Enceladus (GSE) analog 
($0.67 < z_{\rm disrupt} < 3$ and $M_{\rm sub}/M_{\rm MW} > 0.2$) 
and an LMC analog ($V_{\rm \max,sub} > 55$ km s$^{-1}$, $z_{\rm infall} < 0.16$, $30$ kpc $< d_{z=0} < 70$~kpc). The halos are drawn from the
\textsc{c125--1024} parent cosmological simulation \citep{Mao2015}, and 
the suite was resimulated with a particle mass of 
$m_{\rm part} = 4.0 \times 10^5\, M_\odot$. A full description is given in~\cite{buch2024milky}. Table~\ref{tab:mwest}, 
summarizes the host and merger properties. For a complete parameters list, including properties of the LMC and GSE analogs, see Table 2 of~\cite{buch2024milky}.

We also include eight halos selected from the \symphony\ simulation suite \citep{Nadler_2023}, which have masses comparable to the MW, and are simulated with the same resolution as the \mwest\ halos. From the full set of 45 halos, we select halos that have undergone no significant mergers (defined as $M_{\rm sub}/M_{\rm host} > 0.1$) in the past 5~Gyr. These quiescent halos are presented to provide a baseline against 
which to compare the results from the \mwest\ suite. 
We list the IDs and basic properties of these hosts in Table~\ref{tab:symphony}.

% ####fixed table ####
\begin{deluxetable}{l S[table-format=1.2] S[table-format=1.2]
           S[table-format=2.1] S[table-format=1.2] S[table-format=1.2]}
\tablewidth{0pt}

\tabletypesize{\scriptsize}
\tablecaption{Summary of \mwest\ merger and host properties.\label{tab:mwest}}
\tablehead{
\colhead{Halo ID} &
\colhead{$M_{\rm peak,MW}$} &
\colhead{\begin{tabular}{c}$M_{\rm LMC}$/\\$M_{\rm MW}$\end{tabular}} &
\colhead{$d_{\rm peri}$} &
\colhead{$t_{\rm peri}$} &
\colhead{$c/a$} \\
\colhead{} &
\colhead{[$10^{12}\,M_\odot$]} &
\colhead{} &
\colhead{[kpc]} &
\colhead{[Gyr]} &
\colhead{}
}
\startdata
Halo 004      & \snum{1.14} & \snum{0.18} & \snum{13.6} & \snum{0.17} & \snum{0.60} \\
Halo 113      & \snum{1.12} & \snum{0.03} & \snum{44.6} & \snum{0.22} & \snum{0.80} \\
Halo 169      & \snum{1.62} & \snum{0.29} & \snum{58.1} & \snum{-0.36} & \snum{0.92} \\
Halo 170{$^b$} & \snum{1.31} & \snum{0.26} & \snum{8.5}  & \snum{-0.22} & \snum{0.64} \\
Halo 222      & \snum{1.15} & \snum{0.28} & \snum{58.5} & \snum{0.63} & \snum{0.84} \\
Halo 229{$^a$} & \snum{1.78} & \snum{0.01} & \snum{58.7} & \snum{0.25} & \snum{0.64} \\
Halo 282      & \snum{1.35} & \snum{0.06} & \snum{27.2} & \snum{0.61} & \snum{0.70} \\
Halo 327      & \snum{1.20} & \snum{0.14} & \snum{42.7} & \snum{-0.10} & \snum{0.70} \\
Halo 349      & \snum{1.44} & \snum{0.22} & \snum{47.8} & \snum{-1.32} & \snum{0.64} \\
Halo 407      & \snum{1.15} & \snum{0.09} & \snum{53.5} & \snum{-0.01} & \snum{0.78} \\
Halo 659{$^a$} & \snum{1.62} & \snum{0.06} & \snum{36.2} & \snum{-0.60} & \snum{0.75} \\
Halo 666      & \snum{1.58} & \snum{0.46} & \snum{77.4} & \snum{-0.04} & \snum{0.58} \\
Halo 719      & \snum{1.35} & \snum{0.38} & \snum{43.8} & \snum{0.09} & \snum{0.76} \\
Halo 747      & \snum{1.48} & \snum{0.05} & \snum{22.0} & \snum{-0.23} & \snum{0.82} \\
Halo 756      & \snum{1.82} & \snum{0.10} & \snum{66.0} & \snum{-0.02} & \snum{0.82} \\
Halo 788      & \snum{1.70} & \snum{0.04} & \snum{33.8} & \snum{0.00} & \snum{0.52} \\
Halo 975      & \snum{1.17} & \snum{0.29} & \snum{13.2} & \snum{-0.09} & \snum{0.69} \\
Halo 983      & \snum{1.38} & \snum{0.20} & \snum{43.0} & \snum{-0.07} & \snum{0.72} \\
\enddata
\tablenotetext{a}{Had a previous pericentric passage at distances greater than 100~kpc.}
\tablenotetext{b}{Has two close pericentric passages within 2~Gyr; we use the first as the present-day snapshot.}
\tablecomments{Columns show: (1) Halo ID; (2) $M_{\rm peak}$ for the MW host; (3) $M_{\rm LMC}/M_{\rm MW}$;
(4) pericentric distance; (5) time of pericenter relative to present day (0~Gyr; positive values indicate future evolution);
and (6) axis ratio of the host halo prior to infall.}
\end{deluxetable}

%% fixed. 
\begin{deluxetable}{l S[table-format=1.2] S[table-format=1.2]}
\tabletypesize{\scriptsize}
\tablecaption{Summary of \symphony\ host properties.\label{tab:symphony}}
\tablehead{
\colhead{Halo ID} &
\colhead{$M_{\rm peak,MW}$} &
\colhead{$c/a$} \\
\colhead{} &
\colhead{[$10^{12}\,M_\odot$]} &
\colhead{}
}
\startdata
Halo 023 & \snum{1.37} & \snum{0.57} \\
Halo 247 & \snum{1.26} & \snum{0.43} \\
Halo 268 & \snum{1.17} & \snum{0.58} \\
Halo 364 & \snum{1.23} & \snum{0.77} \\
Halo 567 & \snum{1.24} & \snum{0.73} \\
Halo 825 & \snum{1.31} & \snum{0.84} \\
Halo 926 & \snum{1.17} & \snum{0.67} \\
Halo 990 & \snum{1.27} & \snum{0.57} \\
\enddata
\tablecomments{Columns show: (1) Halo ID; (2) $M_{\rm peak}$ for the MW host; (3) axis ratio of the host halo at $z=0$.}
\end{deluxetable}

\begin{figure*}
  \includegraphics[width=\textwidth]{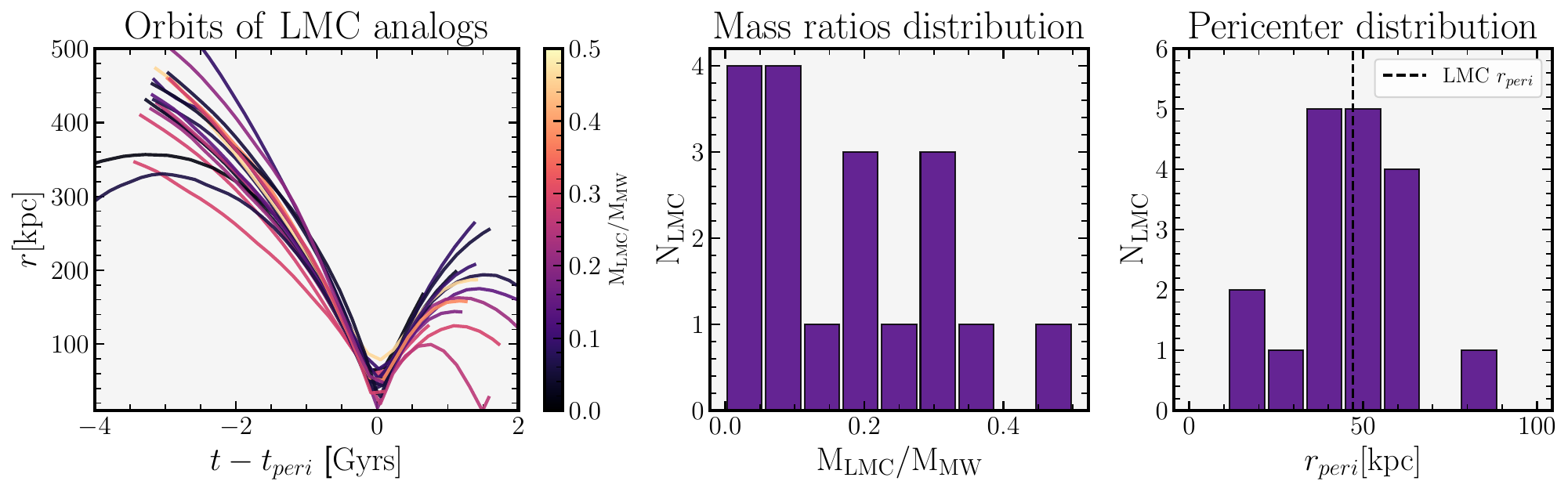}
  \caption{\emph{Left:} Orbit of the LMC analogs for 18 \mwest\ halos included in this analysis are colored by the merger ratio ($M_{\rm LMC}/M_{\rm MW}$). 
  For two of the hosts (229 and 659), the LMC is on its second pericentric passage, having had an earlier 
  pericenter at $r > 100$ kpc in the halo, while for another host, the LMC has not yet reached 
  pericenter by the final snapshot. We have normalized all the times to the pericenter passage 
  of the LMC analog. \emph{Middle:} Histogram of merger ratios between $M_{\rm LMC}/M_{\rm MW}$ 
  for the 18 halos in the \mwest\ suite. The merger ratios range from a ratio of 1:100 at the low 
  end to 1:2 at the high end, with a median ratio of 1:6. \emph{Right:} Distribution of pericentric 
  distances of the LMC analogs. For comparison, the derived pericentric distance for the LMC is 
  shown with the black vertical line.}\label{fig:lmc_orbits}
\end{figure*}

\subsubsection{Defining the Host DM halo}\label{sec:halos}

Particle tracking in both \mwest\ and \symphony\ is done using the \textsc{Symfind} algorithm \citep{mansfield2023symfind}. 
\textsc{Symfind} tracks all halos that a given particle has belonged to, enabling corrections for particle transfers and merger-tree errors. 
%while preventing non-physical particle transfers from hosts to their subhalos or transfers caused by many classes of merger tree errors. 
In many cases, we are interested in identifying the ``smoothly accreted'' mass components of our simulated hosts. Smoothly accreted mass is mass that was directly accreted onto the host without ever belonging to any other subhalo. The mass from all subhalos with $M_{\rm sub} > 1.2 \times 10^8\, M_\odot$ (corresponding to 300 particles) 
are tracked and removed with \textsc{Symfind}, while matter accreted below this limit is considered to belong to the smoothly 
accreted halo. The smoothly accreted mass corresponds to $\approx 30\%$ of the peak mass of the halo in agreement with \cite{Genel10}.
 
\subsubsection{LMC-analog orbits and reference frames}\label{sec:orbits}
We analyze the dynamical response of the host halo to the LMC infall starting at a lookback time of 2~Gyr before the first infall until the present day, with the halo-centric frame and LMC position centered on those determined by the \textsc{Rockstar} halo finder.\footnote{In order to get the LMC at the right distance, several of the \mwest\ hosts were 
run a little over 1~Gyr past $z=0$. To track the halo response as long as possible, we follow each host to its final simulated snapshot.} 

We use the \textsc{Rockstar} halo finder instead of \textsc{Symfind} to measure 
LMC positions because \textsc{Symfind} counts all subhalos whose half-mass radii intersect with their host centers as being disrupted or temporarily disrupted. All particle-tracking subhalo finders must use a comparable technique to remove merged but still bound subhalos from 
the catalog \citep{han_2018,diemer_2023}, although \textsc{Symfind}'s method is relatively aggressive in counting these subhalos as disrupted. \textsc{Symfind}'s criteria led to some LMC analogs being flagged as temporarily disrupted at pericenter. While this has no impact on 
particle associations, we use it to identify the central's smoothly accreted matter. This does suggest that \textsc{Symfind} possibly removes too many high-mass, low-radius halos, as noted 
by the original authors \citep{mansfield2023symfind}. Although \textsc{Rockstar} has problems 
with losing track of subhalos at early times (relatively high $m/m_{\rm peak}$ ratios; \citealp{mansfield2023symfind}), 
all the LMCs in our sample are still tracked by \textsc{Rockstar} at pericenter, meaning that nothing is lost by using this subhalo finder.

The orbits of the LMC analogs are colored by the merger ratios ($M_{\rm LMC}/M_{\rm MW}$) 
are shown in Figure~\ref{fig:lmc_orbits} (\emph{top}). Here, the time is normalized by $t_{\rm peri}$, 
where the pericenter is calculated using linear interpolation of the 3D LMC position as a function of 
the scale factor. The parameter $d_{\rm peri}$ is the minimum distance after interpolation (further discussed in 
Section~\ref{sec:power}), and $t_{\rm peri}$ is the lookback time at the scale at which $d_{\rm peri}$ 
occurs. Two of the halos (\textsc{Halo 229} and \textsc{Halo 659}) in the \mwest\ suite have an early pericenter 
at $z > 0.25$ ($t_{\rm infall} < -3$ Gyr) and $d > 100$ [kpc]. For these halos, we analyze them at 
their second pericenter, making them similar to the model presented in~\cite{vasiliev2023dear} of an 
LMC system on its second pericentric passage. For another halo (\textsc{Halo 788}), the LMC analog has not yet 
reached its first pericenter at the final snapshot, so we set $t_{\rm peri}$ to $z=0$. Lastly, for \textsc{Halo 170}, 
which has had two close-in pericenters ($d < 100$ kpc) in the last 2~Gyr, we set $t_{\rm peri}$ to 
the first pericentric passage. The merger ratios (defined as the mass ratio between the LMC and MW at the time when the LMC was first accreted) of the LMC systems are shown in 
Figure~\ref{fig:lmc_orbits} (\emph{bottom}) and range from 1:60 to 1:2 with a median of 1:6. 

Once the orbits of the LMC-analogs are identified, we rotate all the halos so the satellite's orbit lies 
on the $y-x$ plane as shown in Figure~\ref{fig:bfe-decomposition}. The rotation was done by aligning the 
angular momentum of the orbit with the $\hat{z}-$axis of the halo. This rotation was kept fixed throughout 
the evolution of the halo. The reference frame is centered at every snapshot in the halo cusp identified by 
\textsc{Rockstar}. 

\subsection{A Basis Function Expansion representation of the \mwest\ Dark Matter Halos}\label{sec:bfe}

BFE were first introduced to characterize the gravitational field of galaxies 
in \cite{Clutton-Brock72}. During the subsequent decades, BFE have provided a tool 
to simulate and understand the galaxy dynamics. In particular, BFE have been instrumental
in understanding the dynamics of galactic bars \citep[e.g.,][]{Petersen19, Petersen21}, the satellite--halo
interaction \citep{Weinberg98, Choi09, choiphdt, garavito_19, GC21, Petersen_20, Lilleengen_23}
and the response of galactic disks to internal and external perturbations (\cite{Hunt_25}, Varela et al., Petersen et. al., in prep).
BFE have also been applied to cosmological simulations to capture the time-evolving nature
of halo potentials \citep{Lowing11, Sanders20, arora2022stability}. This is particularly useful for 
reconstructing the orbits of halo tracers \citep{Lowing11, Sanders20, Arora24, arora2024lmc, Brooks25a} such as 
stellar streams \citep{arora2022stability} without the need to re-run a computationally 
expensive simulation. Here, we make use of BFE to quantify and build intuition about the 
dynamical state of cosmological halos. 

We make use of the publicly available code \expcode \citep{EXP} and the associated Python library 
\textsc{pyEXP} to compute and analyze the BFE of the \mwest\ and \symphony\ simulations.  
A comprehensive review of the BFE mathematical background was presented in \cite{Petersen_EXP};
in this section, we briefly summarize the main concepts and equations used to describe the response of the \mwest\ and \symphony\ halos. We work in the spherical coordinate system, which is the natural system to describe halos. Expansions for disk systems are discussed in Section 2.3 in \cite{EXP}.

\subsubsection{BFE representation of the density and potential fields}
A BFE is a complete, orthonormal set of basis functions that 
can uniquely represent any function, given enough terms in the expansion. BFE are useful to represent the density and potential field of galaxies, as they can be used to solve Poisson's equation
for self-gravitating systems. To do this, a set of bi-orthonormal functions, one describing the density $\rho(\textbf{x})$ and 
one the potential $\phi(\textbf{x})$ of the galaxy are used. A basis \textit{set} is then the sum of $\mu$ functions or \textit{modes}, each of which satisfies
Poisson's equation $\nabla^{2}\phi_{\mu}(\textbf{x}) = 4\pi G\varrho_{\mu}(\textbf{x})$, where the functions $\phi_\mu$ and $\rho_\nu$ satisfy the bi-orthogonal condition: 
\begin{equation}
\int_0^\infty \phi_\mu(r) \rho_\nu(r) w(r) \, dr = \delta_{\mu\nu}{\rm ,}
\end{equation}
and where $w(r) \equiv r^2 / 4 \pi G$ is the weighting function, and $\delta_{\mu\nu}$ is the Kronecker delta.

The contributions from each of the basis to the total density and potential are weighted by amplitude coefficients $a_{\mu}$ such that for a discrete system of particles:
\begin{equation}\label{eq:fields}
  \begin{split}
  \rho(\textbf{x}, t) = \sum_{\mu}\,a_{\mu}(t) \rho_{\mu}(\textbf{x}) \\
  \Phi(\textbf{x}, t) = \sum_{\mu}\,a_{\mu}(t) \phi_{\mu}(\textbf{x}),
  \end{split}
\end{equation}
where the coefficients $a_{\mu}$ can be found using the orthonormal properties of $\varrho_{\mu}$ 
and $\phi_{\mu}$: 
\begin{equation}\label{eq:coefficients}
  a_{{\mu}} = \dfrac{1}{N} \sum_{i}^{N} \phi_{\mu} (x_{i}),
\end{equation}
where we have used that all the particles have the same mass, as is the case in the halos of the \mwest\ and \symphony\ simulations.
The Poisson's equation is separable in conic coordinate systems, yielding a one-dimensional ODE for each component. In spherical coordinates, the natural system to describe DM halos, 
the angular function modes $\theta$ and $\phi$ are represented by the spherical 
harmonics $Y_{l}^{m}(\theta, \phi)$ that satisfy the orthogonal conditions needed to build the basis (with $\mu = (n,l,m)$):
\begin{equation}
  \begin{split}
  \rho_{\mu}(\textbf{x}) = \rho_{n}(r) Y_{l,m}(\theta, \phi)\\
  \phi_{\mu}(\textbf{x}) = \phi_{n}(r) Y_{l,m}(\theta, \phi).
  \end{split}
\end{equation}

\begin{figure*}
  \centering
  \includegraphics[width=0.33\textwidth]{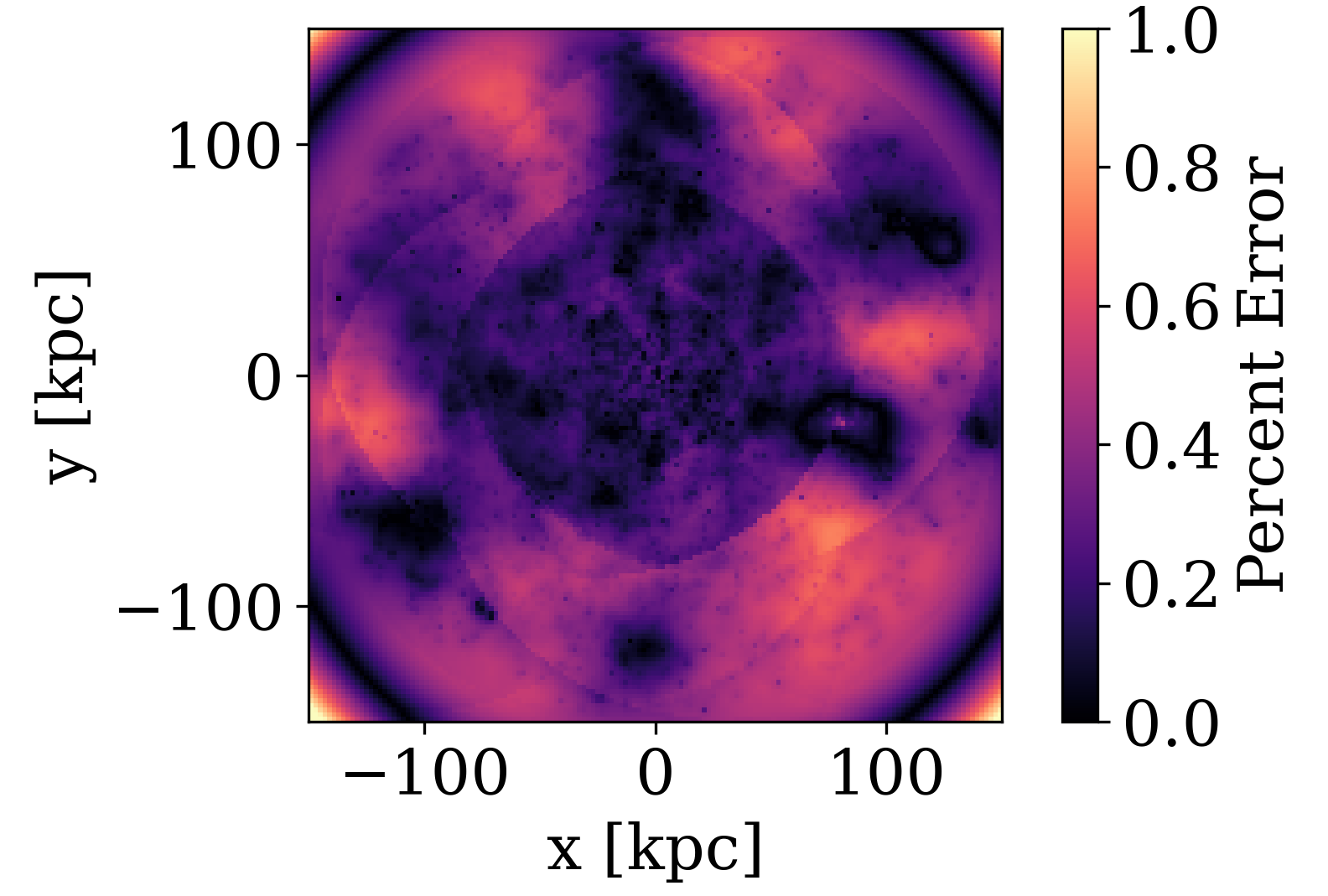}
  \includegraphics[width=0.35\textwidth]{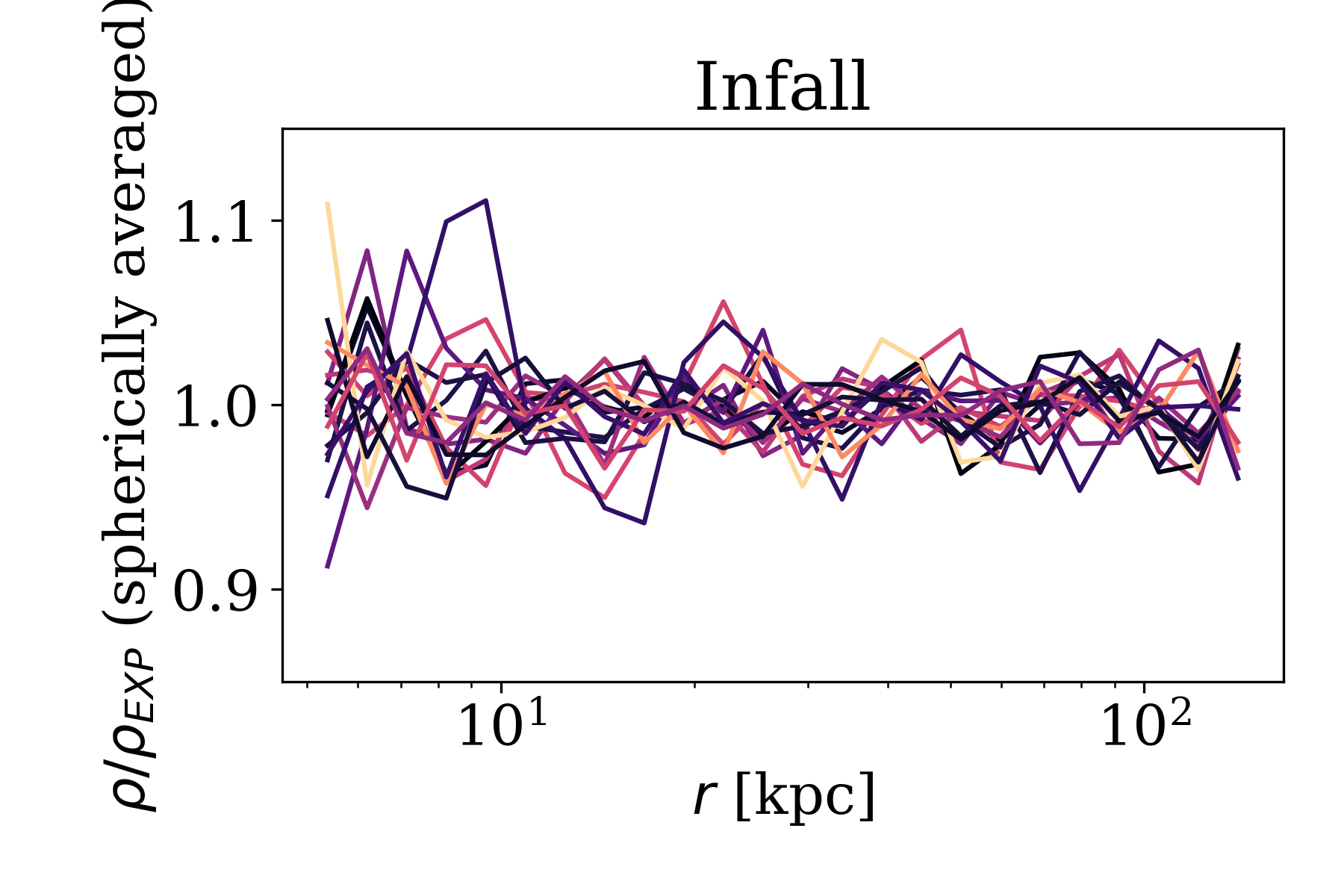}
  \includegraphics[width=0.27\textwidth]{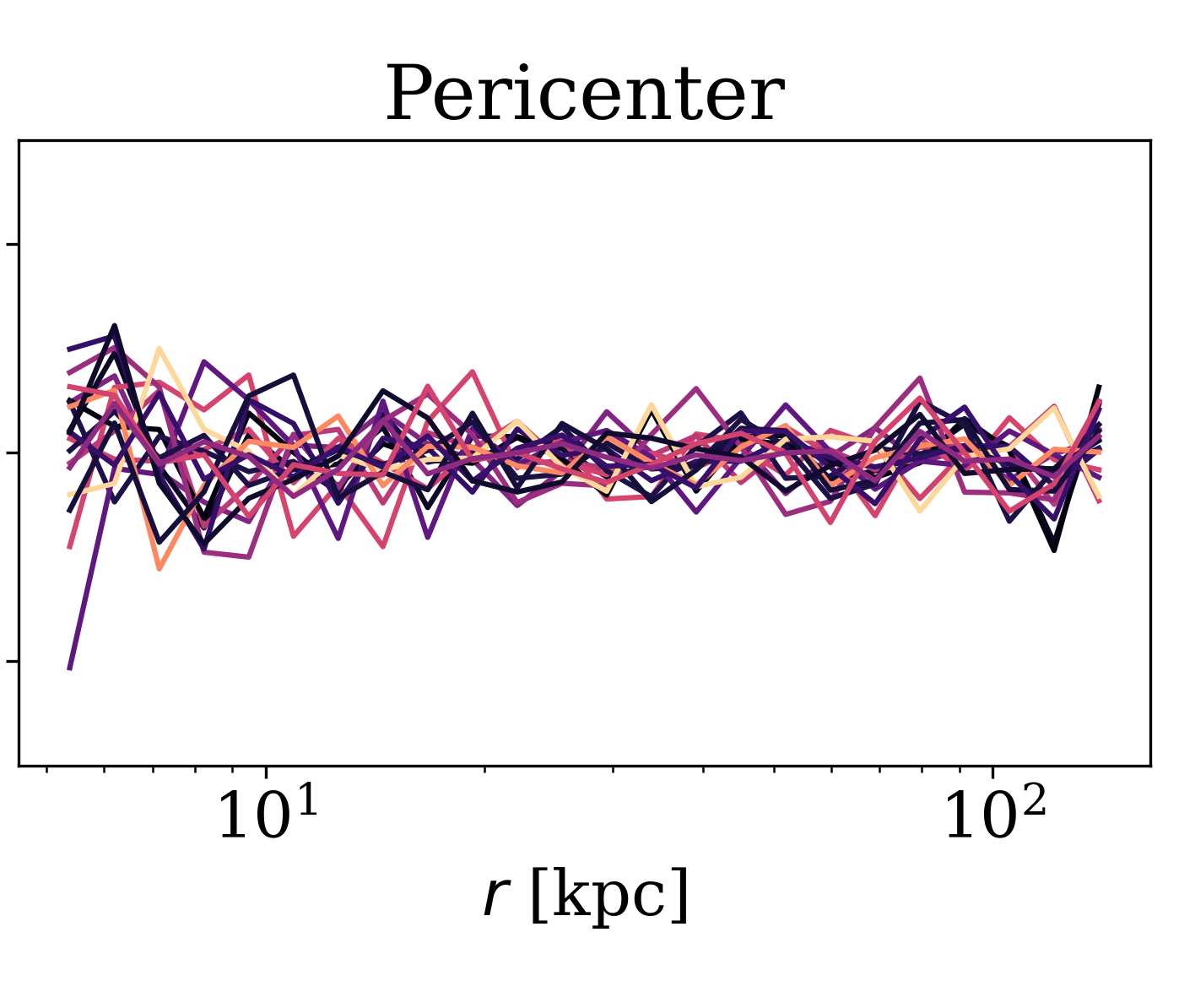}
  \caption{\emph{Left:} Percentage error 
  on the gravitational potential reconstructed using the BFE compared to the potential 
  computed with the tree code method for the raw particle data. The errors across the 
  $z=0$~kpc slice are within 2.5\%. \emph{Right:} Spherically averaged density residuals 
  between the NFW basis and the particle data of all the 18 \mwest\ halos at infall 
  (\textit{center panel}) and at pericenter (\textit{right panel}). Beyond 10~kpc, 
  the residuals are within $\approx 5\%$. In the inner halos, the density computed from 
  the particles are subject to Poisson noise, leading to larger residuals.}\label{fig:dens_residuals}
\end{figure*}

For the radial modes, one needs to build a basis that: $i)$ represents the radial structure of DM halos, $ii)$ is bi-orthogonal for $\phi_{\mu}$ and $\rho_{\mu}$, and $iii)$ satisfies Poisson's equation. 
These three conditions lead to solutions for $\rho_n$ and $\phi_n$ using a generalized form to solve the generalized form of the Poisson equation, the Sturm--Liouville (SL) equation (see Section 2.2 in~\citealt{Weinberg1999}). \expcode solves
the SL problem numerically using the density profile specified by the user. 
The resulting eigenfunctions of the input profile, computed via solution to the SL equation, is the basis that is used to compute using equation~\ref{eq:coefficients}. 

To facilitate straightforward comparison between the 18~\mwest\ halos, we use a common basis for the entire
sample. This choice optimizes interpretability vs precise representation of the density and potential. For the 
present study, we are not seeking to perfectly represent the simulations, but rather to examine dynamical 
evolution indicators --- best accomplished through a unified basis. We use an NFW profile with a scale radius 
$r_s$ of $25$ kpc. This scale radius was chosen based on fits to the particle data across the simulation suite 
and was found to provide a reasonable fit 
to the structure of the dark matter halo across the previous $\sim 10$~Gyr. 
Any deviations from this profile will be adequately represented by the higher-order radial harmonics ($1 < n \leq 10$). The halos are 
expanded between $r_s/100 < r$ [kpc] $< 6\times r_s$ ($0.25$ kpc $< r < 150$ kpc) using a radial 
order of $n_{\rm max} = 10$ and five azimuthal basis functions $l_{\rm max} = m_{\rm max} = 5$. 
In Appendix~\ref{sec:appendix}, we discussed in detail the process of choosing a basis and computing
coefficients with \textsc{pyEXP}. For the purposes of this paper, the order of the expansion 
provides enough resolution to study the general response of the DM halos. That is, 
we can reconstruct the potential and density fields of the halo 
using equations~\ref{eq:fields} within 10\% of that computed with particle data. 

Figure~\ref{fig:dens_residuals} (\emph{left}) shows a comparison of the reconstructed potential 
from the BFE with the potential computed using the raw particle data from a Barnes--Hut tree 
code approximation \citep{Grudic2021}. The BFE accurately represents 
the particle data to the sub-percent level, with the largest errors occurring in regions 
surrounding subhalos with masses below the limit of the particle tracking algorithm (and 
therefore assigned to the smoothly accreted halo). 
The efficient substructure removal provided by the 
\textsc{Symfind} halo finder enables an accurate reconstruction with a low-order expansion.
In the \emph{center} and \emph{right} panels of 
Figure~\ref{fig:dens_residuals} we show the ratio between the particle density ($\rho$) and 
the density from the BFE expansion ($\rho_{\rm EXP}$) for the halo at LMC infall (\emph{center}) 
and pericenter (\emph{right}). Most halos show $0.95 < \rho/\rho_{\rm EXP} < 1.05$ 
with an error primarily dominated by Poisson noise. 

The BFE representation of the potential and density of each halo is fully captured in the $150$\footnote{The length of the expansion is set by $((l_{max}+1)\times l_{max}))/2 \times n_{max}$} coefficients series that can be easily used to query the potential and density at any point in the system. This representation is substantially less computationally 
intensive than the corresponding tree representation, highlighting
the power of BFE to compress dynamical information.

\subsubsection{Using BFE to quantify the contributions from the monopole, dipole, and quadrupole harmonics}
As shown in Figure~\ref{fig:dens_residuals}, BFE are efficient way to compress the information of the
density, potential, and acceleration fields. Another advantage of using BFE is to decompose the 
response of the halos in harmonic modes. A useful way to represent the contribution of each harmonic to the total
response of the halo is by computing the \textit{gravitational potential energy} or \textit{power}. 
The gravitational potential energy ($W$) is the amount of potential energy 
available to do work in the halo. In the BFE formalism, the power in each harmonic is related to the volume integral of the density 
and the potential:
\begin{equation}
  W = -\frac{1}{2} \int \rho(\bf{x}) \Phi(\bf{x}) d\bf{x},
\end{equation}
which can be computed directly from the coefficients of the basis 
making use of its bi-orthonormal properties:

\begin{equation}
  W = \frac{1}{2} \sum_{\mu} \varrho_{\mu} \phi_{\mu} = -\frac{1}{2} \sum_{\mu} a_{\mu}^2 = \sum_{\mu} W_{\mu}.
\end{equation}

As such, one can compute the gravitational energy from each harmonic 
$\mu$ or a set of harmonics. The normalization of the spherical harmonics makes the power in each harmonic rotation-invariant, such that when considering a particular harmonic in totality, the analysis is agnostic as to the orientation or principal axes of the galaxies.
In this work, we are particularly interested 
in the departure of our halos from spherical shapes, so we concentrate on examining the different $l$ harmonics, summed over their radial ($n$) and azimuthal ($m$) contributions.
As such, the gravitational power in each $l$ harmonic, using:
\begin{equation}
  W_{l} = -\frac{1}{2} \sum_{n, m} a^2_{n, l, m}.
\end{equation}

\subsubsection{Physical interpretation of a BFE decomposition.}

To explore the response of the smoothly accreted halo to the infall of the LMC, we mainly focus
on the gravitational power in the first- and second-order $l$ harmonics of the BFE. For a
spherical basis, the various $l$ orders correspond to the spherical harmonics, the $l=1$ harmonic measures
the dipole response of the halo, while $l=2$ measures the quadrupole response. As shown 
in \cite{Petersen_20}, \cite{GC21}, and \cite{Lilleengen_23}, we expect 
these harmonics to show the largest excitation in response to the LMC infall. Furthermore, we 
expect the COM displacement discussed in \cite{GC21} to manifest as 
a strong dipole harmonic in the plane of the satellite orbit and opposite the position of the LMC. This dipolar response has been found to be strong in 
cosmological halos and plays a major role in the shape of galactic disks \citep{Gomez16}.

Similarly, the triaxial shape of the halo should be captured by the even $l$ and $m$ harmonics. In particular, as the lowest-order even angular harmonic, the quadrupole will be the strongest harmonic in representing the triaxiality of the halo. The dynamical friction wake, on the other hand, is expected to show a more complex 
structure than in idealized $N$-body simulations, but should be present in the quadrupole and higher-order harmonics 
 \citep{GC21}. In Section~\ref{sec:classical_wake}, we expand 
our analysis to the higher-order harmonics in order to fully characterize the dynamical friction 
wake.

\subsubsection{Halo shape measurements}~\label{sec:inertia_tensor}

We characterize the halo shapes in each halo by finding the ellipsoid that best fits 
the halo density within a shell defined within $50-150$~kpc \footnote{Note that the halo shape within the inner 50~kpc can defer to that in the outskirts presented here.}. 
The ellipsoid is characterized by three \textit{principal axes} (PA)
($\vec{a}, \vec{b}, \vec{c}$) in 
the halocentric coordinate frame as defined in Section~\ref{sec:orbits}. 
We do this at every snapshot of the simulation from redshift $z=1$ to present-day.
We find the ellipsoid PA by diagonalizing the moment of inertia tensor
defined as: 
\begin{equation}\label{eq:I}
  I_{ij} = \dfrac{1}{M_{shell}} \sum_{i,j}^{3} {m_p x_{i} x_{j}}, 
\end{equation}
where $i, j$ are the coordinates of the particles of the halo in halo-centric coordinates, 
$m_p$ is the particle mass, and $M_{shell}$ is the total mass inside $50-150$~kpc.
The eigenvectors and the square root of the eigenvalues of the inertia tensor correspond 
to the directions and magnitudes of the PA. In our notation, the PA always satisfies 
$a \geq b \geq c$. For a spherical halo $a = b = c$, for a
prolate halo $a \geq b = c$, and for an oblate halo $a = b \geq c$. The orientation 
of the axis is quantified by the angle $\theta$ between the larger PA $a$ and the $\hat{x}$
axis of the halo. 

In the analysis presented in Section~\ref{sec:results}, we report results in terms of 
the axis ratios $c/a$, $b/a$, $\theta$, and the triaxiality parameter ($T$), commonly used to summarize the halo shape. $T$ is defined as: 
\begin{equation}
  T = \frac{a^2 - b^2}{a^2 - c^2},
\end{equation}
where $T$ is zero for a perfect oblate halo, and $T$ is unity for a perfect prolate halo 
\citep{Warren1992}. The transition from oblate to prolate 
takes place around $1/3 \leq T \leq 2/3$ where halos are \textit{triaxial} \citep{Warren1992}. 

Since we are mainly interested in global measurements of the halo shape, we use the simple definition of the inertia tensor of equation~\ref{eq:I}. Even though more sophisticated methods to find halo shapes have been discussed in the literature \citep[e.g.,][]{Thob19}.

We compute the halo shape with the \texttt{halo\_shape} 
function in \texttt{pynbody} using a single homeoidal shell between $50<r$ [kpc] $< 150$ 
as a proxy for the halo shape in the radial range of interest. This is
roughly the range where previous simulations and observations predict a strong halo response to the infalling LMC. All the quantities describing the halo structure, such as axis ratios $b/a$, $c/a$, and triaxiality $T$, are computed as the median between $-3\,$ Gyr $< t-t_{\rm peri} < -1\,$ Gyr, where $t_{peri}$ is the pericenter of the satellite. We adopt a similar timescale for the \symphony\ halos.

\section{Results}~\label{sec:results}

\begin{figure*}
\includegraphics[width=\textwidth]{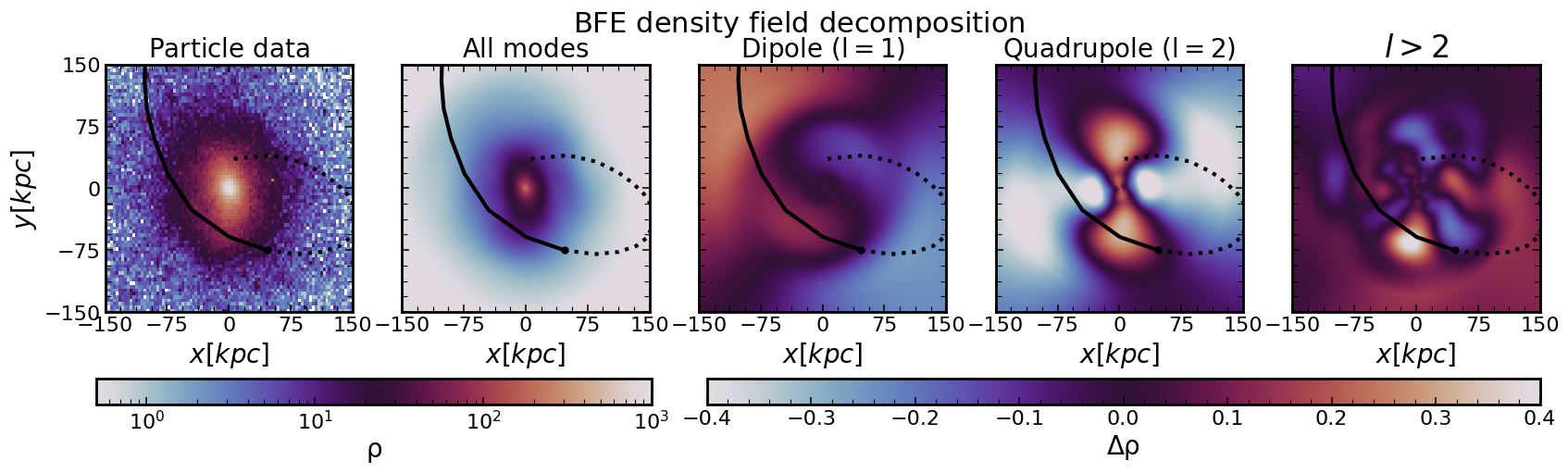}
  \caption{Example density field computation using the BFE and decomposition for \textsc{Halo 349} from 
  the \mwest\ simulation suite, showing from left to right (1) projected density of 
  smoothly accreted host particles, (2) full BFE with $l=5$ and $n=10$, (3) density of $l=1$ 
  harmonic relative to $l=0$, (4) density of $l=2$ harmonic relative to $l=0$, 
  and (5) density of $l>2$ harmonics relative to $l=0$. For the leftmost plot, the particle density 
  is projected in the X-Y plane. The plot 
  shows a cross-sectional slice at $z=0$. The colorbars for panels (1) and (2) are 
  scaled arbitrarily, while plots (3-5) show the density contrast. 
  The dashed line shows the orbit of the LMC analog, and the black dot shows the current position 
  (chosen to be roughly at the pericentric passage).}
  \label{fig:bfe-decomposition}
\end{figure*}

Halos are made out of harmonic modes, but what are those modes, and how 
much do they contribute to the density field of the halo? 

Figure~\ref{fig:bfe-decomposition} shows the density field of \textsc{Halo 407} at the time of the satellite's pericentric 
passage decomposed into its harmonic modes. The first and second columns show the full 
density, computed with the particle data and the full expansion. Subsequent columns show contributions from the $l=1$ harmonic (\emph{3rd column}), 
the quadrupole ($l=2$) harmonic (\emph{4th column}), and all higher-order ($l>2$) harmonics 
(\emph{5th column}). The $l=1$ harmonic largely exhibits overdensities in one hemisphere of the halo ($z > 0$ and $y > 0$), possibly 
due to the halo density center displacement induced by the satellite. In contrast, 
the $l=2$ harmonic represents the elongated shape of the halo, mainly along the $y$-axis (see Section~\ref{sec:correlations} for a detailed analysis on this). 
All harmonics contribute to the overdensity of the dynamical-friction wake trailing the satellite; as the number of harmonics increases,
the wake is resolved more finely \citep{GC21, Simon_22}.

\begin{figure*}
  \centering
  \includegraphics[width=1.8\columnwidth]{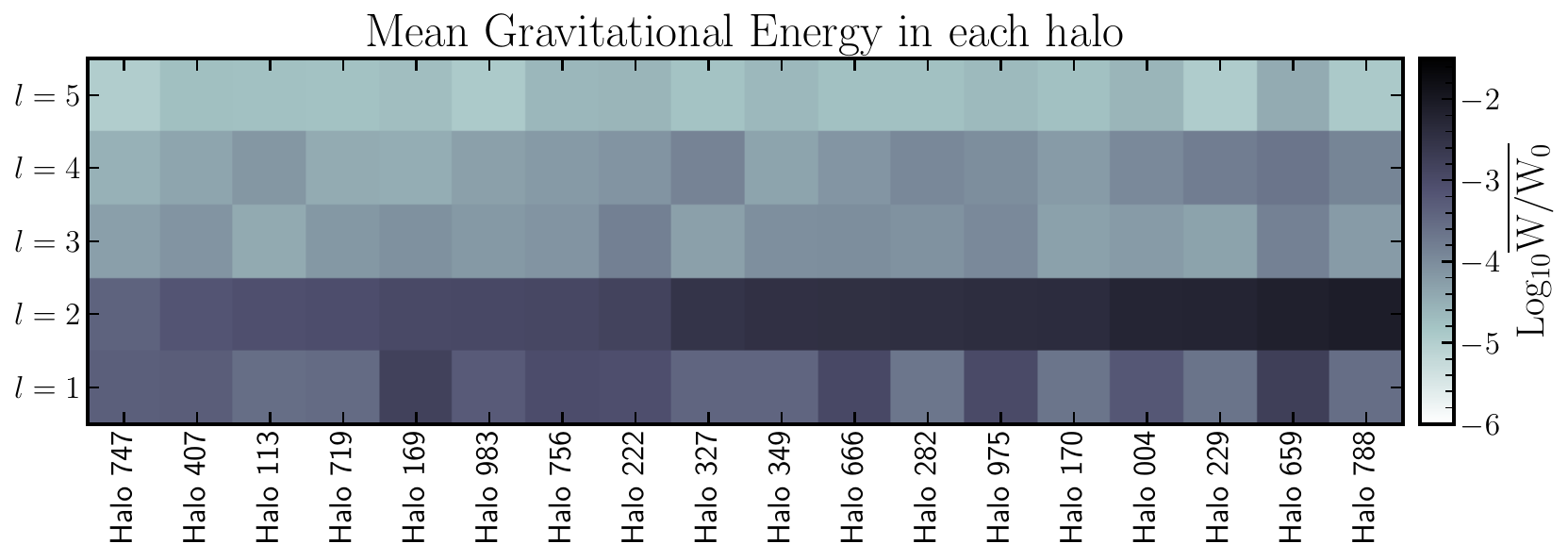}
  \caption{\textit{Left:} Mean gravitational power relative to the monpole ($W_0$) through the evolution of each halo for the $l=1$--5 modes. On average, the quadrupole ($l=2$) and dipole ($l=1$) harmonics are the second and third most dominant modes after the monopole.}
  \label{fig:all-modes}
\end{figure*}

The structure of halos can be fully described as a linear combination of their harmonic modes. As shown in Figure~\ref{fig:bfe-decomposition}, some harmonics have larger amplitudes than others. To build intuition about which harmonics dominate the description of halos, we show all modes present in the \mwest\ suite in Figure~\ref{fig:all-modes}.
We show the amplitude of the gravitational energy 
in each harmonic, averaged over the last $\approx4$~Gyr of evolution of each halo in the \mwest\ suite. For all halos, 
most of the gravitational energy is in the monopole and in the quadrupole, followed by the dipole. Higher-order harmonics, such as the $l=5$, contain little
gravitational power, and we do not consider them further for our purposes of characterizing the broad halo response. Nevertheless, as shown in the left panel of Figure~\ref{fig:dens_residuals}, 
 an $l=5$ pattern is visible in the residuals. Thus, to 
achieve higher precision in the density, potential, or acceleration fields, 
one can increase the order of the expansion. 

\begin{figure}[ht]
  \centering
  \includegraphics[width=1.0\columnwidth]{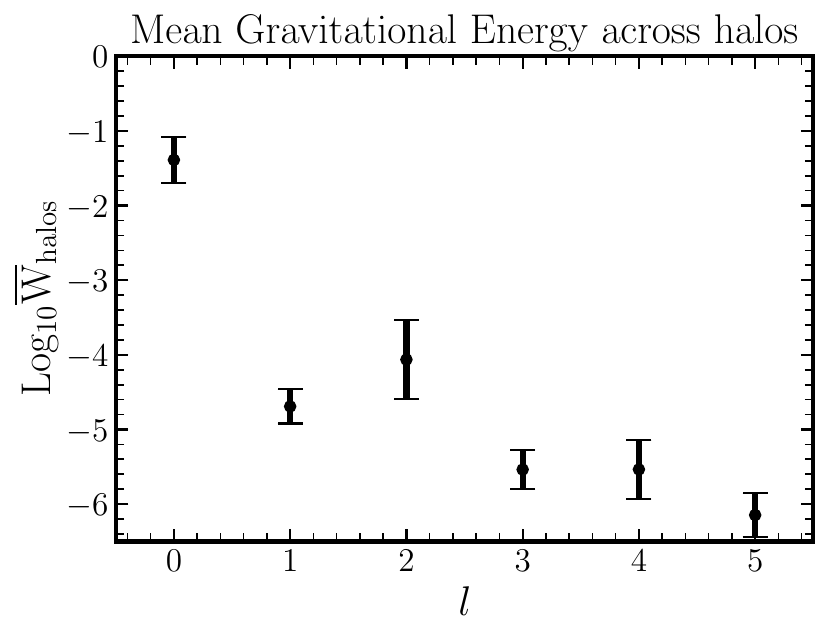}
  \caption{Mean gravitational power averaged over the \mwest\ suite for each harmonic mode. Error bars show the standard deviation across all halos over their evolution. DM halos are mainly characterized by the monopole, quadrupole, and dipole terms. Higher-order modes have lower amplitude and may not affect the halo as a whole, but they contain information about smaller-scale perturbations.}
  \label{fig:all-mean-modes}
\end{figure}

The mean gravitational energy across all halos is plotted in Figure~\ref{fig:all-mean-modes}.
The error bars represent the standard deviation across all 18 halos in the \mwest\ halos.
On average, the monopole ($l=0$) is at least two orders of magnitude larger than the 
quadrupole ($l=2$) and three orders of magnitude larger than the dipole ($l=1$). Since the monopole was chosen to 
follow the NFW halo (see Section~\ref{sec:bfe}) that best fitted all the halos, it is expected that it has the largest 
amplitude. Each of these harmonics is correlated with a physical quantity of the halo, 
this will be discussed in the next subsections. 

In the following subsections, we quantify the evolution of the halo response to the infall 
of LMC-like satellites in terms of the amplitudes of the modal response characterized by the 
coefficients of the BFE. 

\begin{figure*}[ht]
  \includegraphics[width=0.99\textwidth]{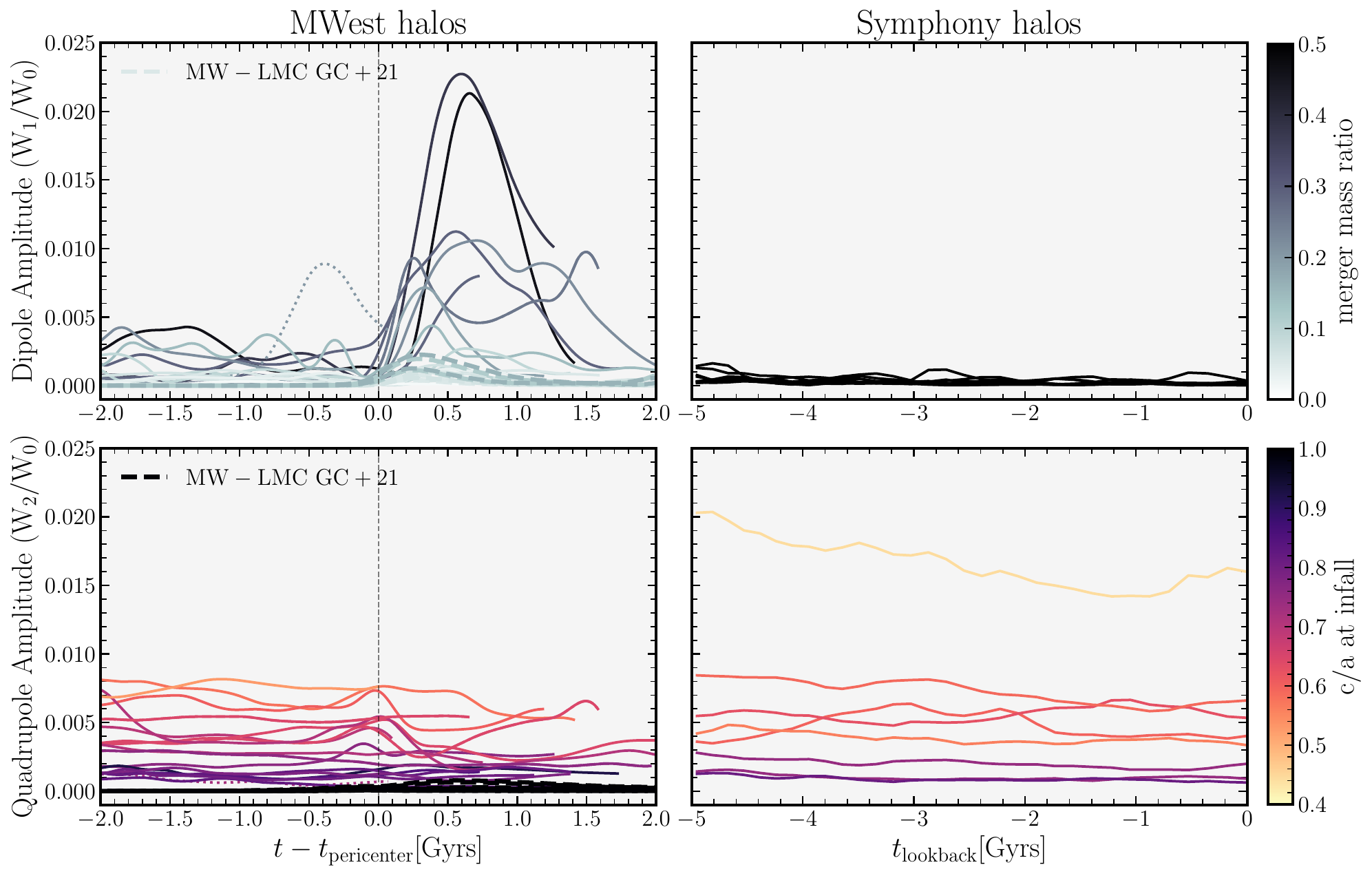}
  \caption{\emph{Top:} Relative power in the $l=1/l=0$ harmonics as a function of time for the 18 \mwest{} hosts (\emph{left} panels), compared with 8 quiescent hosts from the \symphony{} suite (\emph{right} panels) and with the 8 idealized MW–LMC merger of \cite{GC21} (dashed lines). Lines are colored by the median merger mass ratio, averaged over ~2 Gyr prior to infall ($-3 ,\mathrm{Gyr} < t - t_{\rm peri} < -1 ,\mathrm{Gyr}$), to highlight the correlation with amplitude of the power in the dipole ($l=1$). The $x$-axis is shifted such that $t-t_{\rm peri} = 0$ corresponds to the time of the LMC’s first pericenter. For hosts in which the LMC has not yet reached first pericenter, $t-t_{\rm peri} = 0$ is defined as the final snapshot. The dotted line shows Halo 170, which experienced an earlier satellite pericentric passage. The dashed vertical line marks $t-t_{\rm peri} = 0$.\emph{Bottom}: Temporal evolution of the relative power in the quadrupole ($l=2/l=0$). Lines are colored by the median axis ratio ($c/a$) to illustrate the correlation with quadrupole strength. Two halos are excluded from this analysis: Halo~788, which has not yet reached pericenter, and Halo 983, which experienced a massive merger immediately prior to the LMC infall. Although not noticeable in the figure, the quadrupoles in idealized simulations peak before the pericenter, but their amplitudes are $\approx$20 times weaker.} 
  \label{fig:power}
\end{figure*}

\subsection{Signatures of halo response in low-order modes}\label{sec:halo_response} 

\begin{figure*}[ht]
  \includegraphics[width=\textwidth]{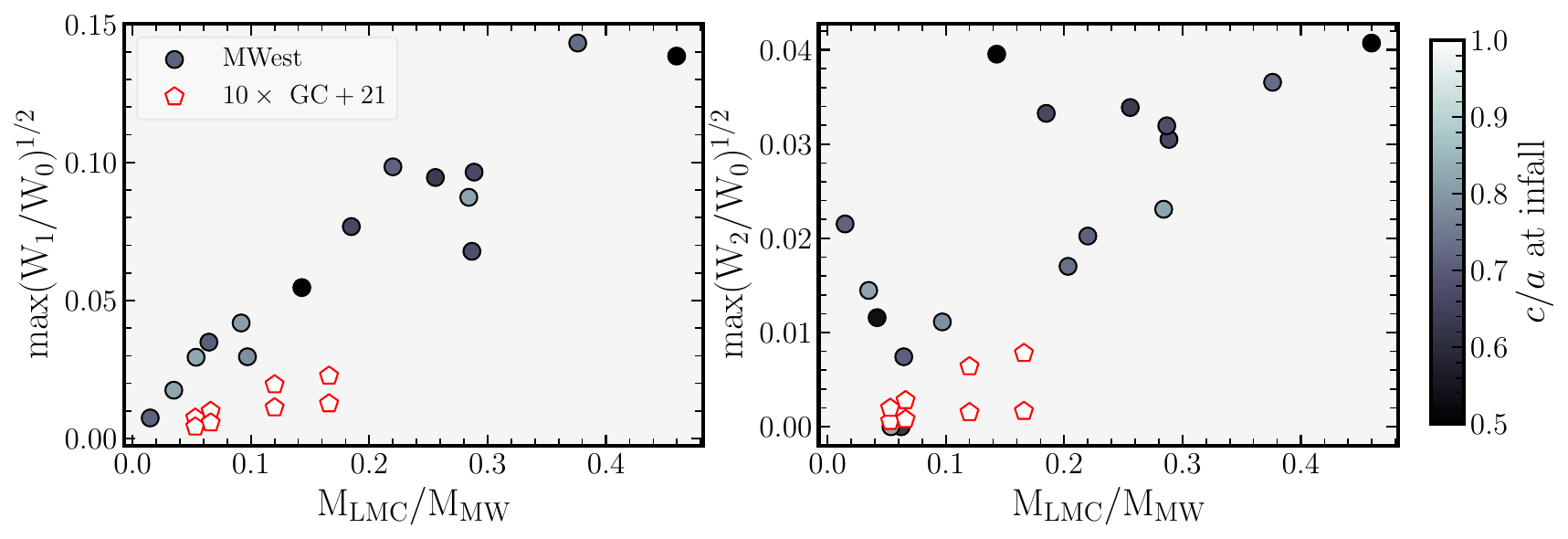}
  \caption{Correlation between maximum power 
  and merger ratio for $l=1$ (\emph{left}) and $l=2$ (\emph{right}). The $l=1$ power is taken as the maximum between $0$~Gyr $< t-t_{\rm peri} < 1$~Gyr, while the $l=2$ power is the maximum between $|t-t_{\rm peri}| < 0.5$~Gyr. Also shown are the $l=1$ and $l=2$ values of power (multiplied by 10) from the 8 idealized simulations in \cite{GC21}. Points are colored by the halo's $c/a$ axis ratio measured when the LMC was just outside the host halo.}\label{fig:strength}
\end{figure*}

\subsubsection{Description of the time evolution}

Figure~\ref{fig:power} (\emph{top}) shows the temporal evolution of the gravitational
energy in the dipole ($l=1$) and quadrupole ($l=2$) harmonics for the 18 halos in the 
\mwest\ (left panels) and the 8 halos in the \symphony\ (right panels) suites. The 
amplitudes of the $l=1$ and $l=2$ harmonics are normalized by the amplitude in the 
monopole ($l=0$) and centered on the time of the first pericenter of the LMC 
($t-t_{\rm peri} = 0$). The color of the lines corresponds to the mass ratio of the 
satellite and host in the dipoles panels (top) and the axis ratios of the halos. 
For comparison, we also show the relative amplitude of gravitational energy of 
the $l=1$ and $l=2$ of the 8 idealized $N$-body MW--LMC simulations in \cite{GC21} halos. 

The \symphony\ halos (absent of LMC-like mergers) show a similar range of power in $l=2$ to the \mwest\ halos, 
but negligible power in $l=1$, consistent with the picture that quadrupole power is 
set early in the halos' assembly history and persists over long time scales \cite{Arora25}, 
while the power in the dipole is induced by perturbations in the halo, such as mergers, 
and predicted to last for several dynamical times \citep{Weinberg_23}. 

The LMC infall induces a strong dipole and weaker quadrupole response in the majority 
of halos in the simulation suite, confirming the results found in idealized $N$-body 
simulations \citep[e.g,][]{Petersen_20, GC21, Lilleengen_23}, and in cosmological 
simulations \citep{Gomez16, Arora25}. For a number of the halos, a strong $l=2$ harmonic 
is already present prior to the first pericentric passage of the LMC (Figure \ref{fig:power}, 
\emph{bottom left}), which correlates strongly with the axis ratio of the halo (illustrated by 
the color bar and computed as the median axis ratio between $-3$ Gyr $< t-t_{\rm peri} < -1$ Gyr). 
In these halos, the LMC response manifests as a small perturbation on top of the preexisting 
quadrupole consistent with \cite{Arora25}. Meanwhile, the dipole power tends to be negligible prior to the first pericentric 
passage and displays a strong peak after the pericentric passage (the one halo that peaks 
prior to LMC pericenter has another major merger that reaches the pericenter at approximately 
$t-t_{\rm peri} = -1$). The dipole response peaks between 0.5--1Gyr after the first LMC 
pericenter, while the quadrupole response peaks at or very near pericenter (around when 
we expect a peak in the dynamical friction wake, indicating a correlation between the 
quadrupole and the wake). The length of the response is on the order of 1--2~Gyr for 
the dipole and 0.5--1~Gyr for the quadruple.

We also show the relative power in the $l=1$ and $l=2$ harmonics induced in the idealized 
MW--LMC simulation from \citep{GC21}. The host halo in this simulation was initialized as 
an idealized Hernquist profile, and the mass ratio of the MW and LMC is $\approx 0.12$. We 
see that, on average, the idealized simulation shows $20\times$ lower power in both the $l=1$ 
and $l=2$ harmonics relative to the monopole.\footnote{The idealized simulations do show an 
enhancement in $l=2$ at pericenter, however, the power is so low that it cannot be seen due 
to the scale in Figure \ref{fig:power}, \emph{right}.} Further analysis is required in order 
to understand the difference in the amplitudes between the cosmological and the idealized halos.

In idealized simulations of isolated galaxies, it has been found that dipoles are the 
easiest modes to perturb in a halo \citep{Weinberg_23}. Dipoles are weakly damped 
harmonics whose lifetime could exceed the age of the Universe. Once a dipole is excited 
(even mildly), it could be supported by intrinsic properties of the halo and provide 
long-term influences on disk evolution observed as lopsidedness \citep{Varela23}.
In cosmological simulations, where a halo is dynamically perturbed by many processes, 
the lifespan of dipoles has not been quantified systematically, but in individual halos they can last for several Gyrs \citep{Gomez16}. 

\subsubsection{Physical Interpretation}\label{sec:correlations}

\begin{figure}
  \includegraphics[width=\columnwidth]{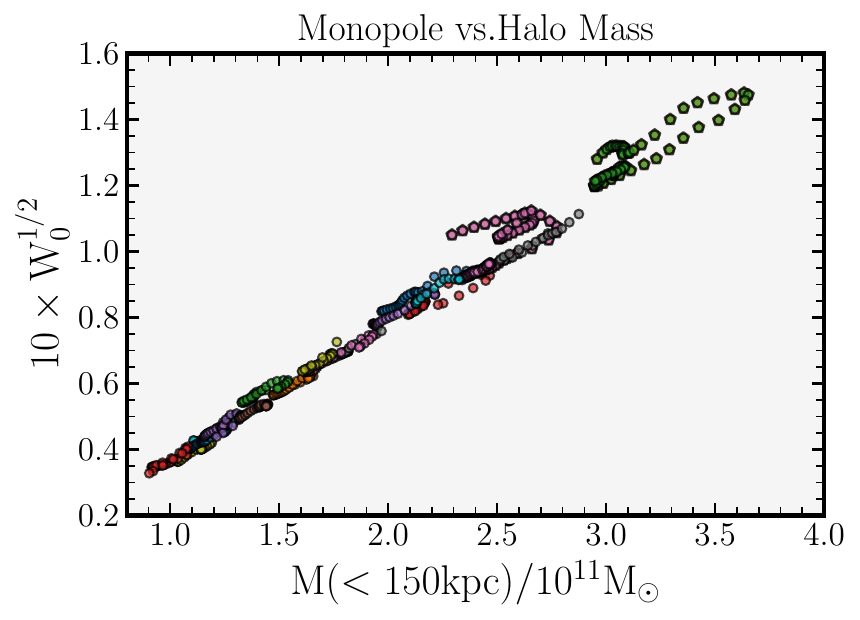}
  \includegraphics[width=\columnwidth]{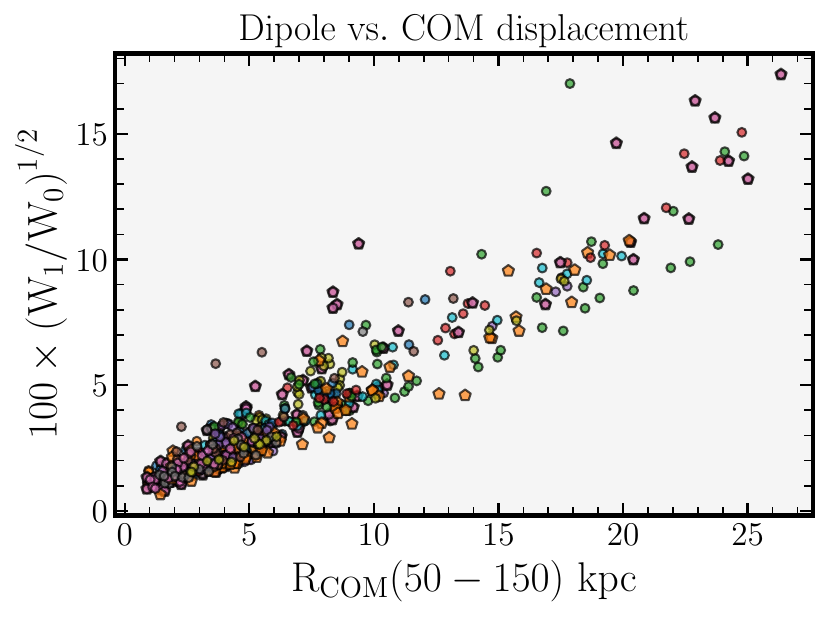}
  \includegraphics[width=\columnwidth]{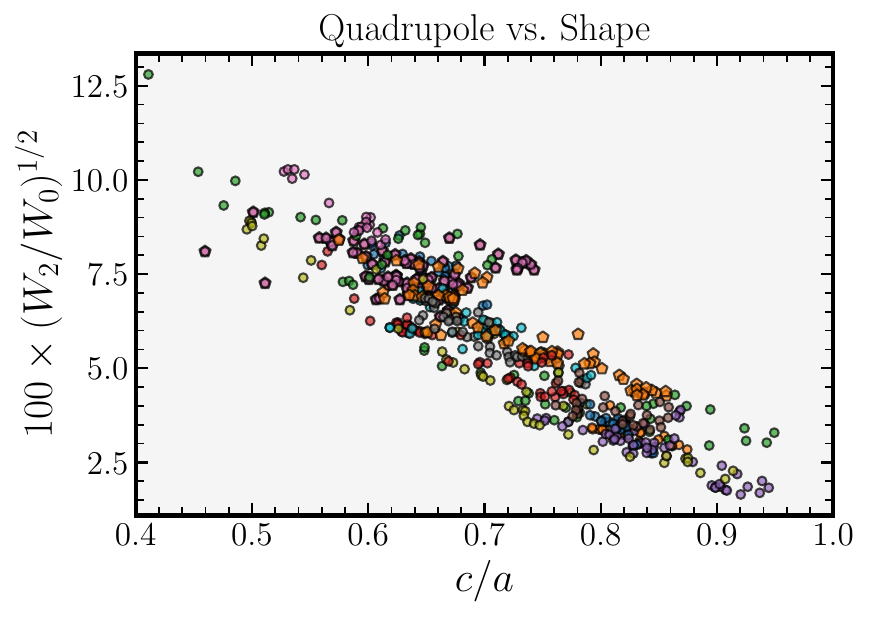}
\caption{Halos' structure correlations with low-order harmonics in the \mwest\ halos. Top panel: 
Correlation of the enclosed smoothly accreted halo mass (within 150~kpc) with the square root of the 
gravitational energy of the monopole ($l=0$). Each dot corresponds to a halo at a given time; each color 
corresponds to the snapshots of the same halo. The one-to-one relationship demonstrates that the monopole 
term always traces the enclosed mass of the halo. Middle Panel: Correlation between the square root of the 
gravitational energy in the dipole and the COM of the outer halo ($\leq$50~kpc). Larger amplitudes in the 
dipoles arise when the halos have experienced larger COM offsets. Bottom Panel: Correlation between the 
square root of the gravitational energy in the quadrupole and the axis ratios of the halos. Larger 
amplitudes of the quadrupole typically represent the most elongated halos.}\label{fig:correlations}
\end{figure}

We now show that the low-order harmonics ($l = 0, 1, 2$) correlate with physical quantities 
and dynamical processes in the halo. We begin by exploring the correlation between the peak 
amplitudes of the dipole and quadrupole and the mass ratio of the LMC to the MW \citep[see also][]{Gomez15}.

In Figure~\ref{fig:strength}, the left-hand panel illustrates a strong correlation between 
the peak amplitude of the dipole and the mass ratio of the LMC to the MW. Notably, this 
correlation is consistent with the results from idealized simulations (indicated by open 
red pentagon markers). These results suggest that measuring the density dipole in the MW 
response will constrain the mass ratio of the MW and the LMC.

The right-hand panel of Figure~\ref{fig:strength} shows the correlation between the peak 
amplitude of the quadrupole and the halo mass ratio. Although there is noticeable scatter, 
a positive correlation is present, indicating that the LMC does induce a quadrupole in the 
MW. However, the magnitude quadrupole peak is almost a factor of two lower than in the 
idealized simulations. Unlike the dipole, the mass ratio alone does not uniquely determine 
the quadrupole's amplitude. As shown in \cite{Arora25}, there are two harmonic modes contributing 
to the total quadrupole: a preexisting quadrupole, which reflects the MW’s intrinsic triaxial 
shape, and an induced quadrupole resulting from the LMC's perturbation. Consequently, the 
total quadrupole is the linear combination of both quadrupoles. Disentangling both signals 
is therefore needed to measure the underlying shape of the MW's halo.

Figure~\ref{fig:correlations} shows three correlations found in the \mwest\ halos. Each halo 
is represented by the same marker color, and each marker represents the values at a given 
snapshot. The top panel shows the correlation between the square root of the gravitational 
energy in the monopole $W_0$ and the enclosed mass of the halo. We measured the enclosed mass 
within 150~kpc because that is the maximum radius used to compute the basis in each halo. 
The linear correlation is expected, as only the monopole contributes to the mass of the halo.

The correlation between dipole motion and the displacement of the density center is shown in the middle panel 
of Figure~\ref{fig:correlations}. The density center displacement of the halo is calculated by measuring 
the distance from the peak density of the host halo to the COM of all particles within the 
50-150\,kpc range. As mentioned in the previous section, while the amplitude of the dipole is 
generally smaller compared to the quadrupole during most of the halo's evolutionary stages, 
it peaks immediately following a satellite's pericentric passage. It is at these moments that 
the halo experiences the most significant disturbance from the satellite, primarily evident 
in the halo's sloshing, or center of mass movement. This phenomenon occurs because the inner 
halo has shorter dynamical timescales than the satellite's orbital period, allowing it to 
react swiftly to the satellite's gravitational force. In contrast, the outer halo, with 
its longer dynamical timescales, shows a delayed response compared to the inner halo. This 
ongoing sloshing between the inner and outer halo continues until the halo reaches a 
relaxed state.

Unlike the monopole--halo mass correlation, there is a noticeable scatter in the dipole--COM 
displacement correlation. The nature of this scatter is non-trivial to quantify, but understanding 
the exact dependence of the halo's dipole with the halo structure is key to characterizing the 
properties of the DM halo response and nature. Future work will explore this relationship. It 
is worth noting that the dipole is also sensitive to the pericentric distance of the satellite, 
the halo shape of the host, the density profile of the halo, and kinematics. In addition, as 
we showed in the middle panel of Figure~\ref{fig:bfe-decomposition}, the dipole tracks part 
of the overdensities in the dynamical friction wake. Even though the dipole--COM displacement 
correlation is a simplified view of the dynamical information that contains the dipole, it is 
a useful correlation to compare across halos and even DM models. 

The quadrupole--shape correlation is shown in the lower panel of Figure~\ref{fig:correlations}.
Geometrically, the shape of a halo represented by the axis ratio ($c/a$) is an axisymmetric 
elongation of the halo in a preferred direction. The quadrupole modes capture this elongation 
as it adds mass in two lobes in the halo aligned with the principal axis of the halo and subtracts 
from the other lobes. The scatter in this correlation is also non-trivial to understand, but most 
likely is mediated by the response of the halo to the satellite and by environmental effects. 
For example, the impact of filamentary accretion is captured mainly by the quadrupole as shown 
in \citep{Arora25}.

Other higher-order harmonics have information and therefore correlate to the halo shape, COM 
displacement, wakes, etc. Decomposing the contribution from each of these harmonics to a 
particular response of the halo is non-trivial. However, recent spectral techniques such as 
multichannel singular spectral analysis (mSSA) have been successfully applied to BFE, illustrating 
that this is possible and powerful to understand the dynamical response of DM halos \citep[e.g.,][]
{Weinberg_21, Arora25, Hunt_25}.

\subsection{Orientation of the Halo Response} \label{sec:orientation}

\begin{figure}[ht]
  \includegraphics[width=1.0\columnwidth]{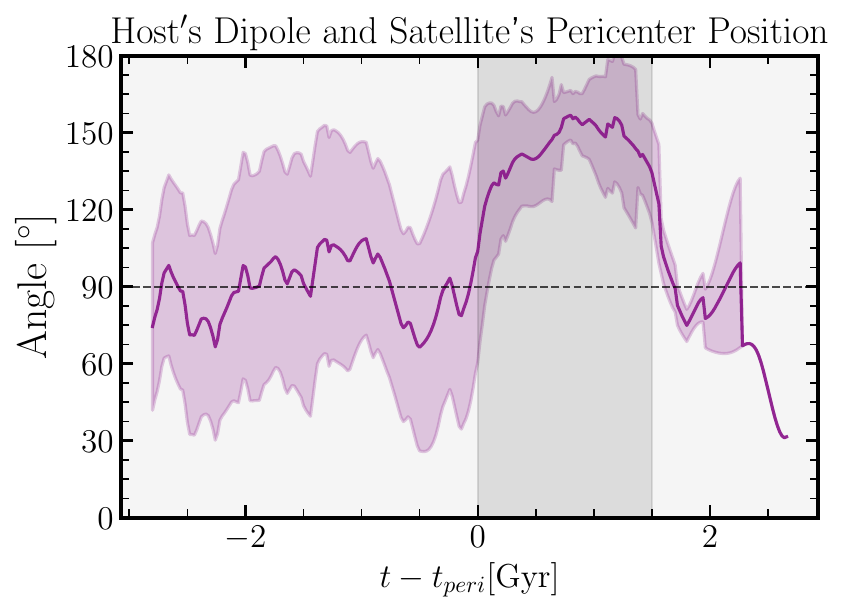}
  \caption{Separation angle between the dipole and the pericenter 
  position of the LMC for all 18 \mwest\ halos as a function of $t-t_{\rm peri}$. We see that 
  the dipole response is anti-aligned with the pericenter position of the halo and in the plane 
  of the satellite (90 degrees with respect to the angular momentum) as a results of angular and linear momentum conservation. This alignment lasts about 
  as long as the peak in the $l=1$ power (\emph{grey shaded region}, as shown in Figure~\ref{fig:power}).}
  \label{fig:orientation}
\end{figure}

As illustrated in Figure~\ref{fig:power}, the amplitude of the response of the halo changes in time. Similarly, the direction (the phase) of the harmonics also changes in time as the satellite orbits around the host. The phase of the harmonics can be characterized at each time step by taking the ratios between the $m$ harmonics for every $l$ mode. In this section, we analyze the phase of the $l=1$ modes and correlate it with the satellite orbit. 

In our halo-centric reference frame, where the halos have been rotated so that the orbits of the LMC-analogs lie on the $x-y$ plane, the $(l=1,m=0)$ coefficient corresponds to the power in the $z$ direction, while the $(l=1,m = \pm 1)$ coefficient corresponds to the power in the $x$ and $y$ directions. This is
\begin{equation}
\begin{split}
  \theta_{dipole} = \mathrm{arccos}\left(\frac{a_{0, 1, 0}}{ a_{0, 1}}\right) \\ 
  \phi_{dipole} = \mathrm{arctan2}\left(\frac{a_{0, 1, 1}}{a_{0, 1, -1}}\right),
\end{split}
\end{equation}\label{eq:dipole_angles}
where the order of the sub-indices follows the notation $a_{n,l,m}$ and
\begin{equation}
a_{0,1} = \left(\sum_{m=-1}^{1}{a_{0, 1, m}^2}\right)^{1/2}. 
\end{equation}

We focus on the $n=0$ harmonics since these harmonics have most of the gravitational energy. They also correspond to the largest scales, and hence are likely to capture the overall halo orientation rather than local deviations. To physically interpret the evolution of the orientation of the halo response, we measure the angles between the dipole ($l=1$) mode and the LMC pericentric position and the LMC's angular momentum vector (shown in the middle column in Figure~\ref{fig:bfe-decomposition}).

If the dipole ($l=1$) is dominated by the satellite response, we expect it to be in the plane of the LMC's orbit (perpendicular to the angular momentum of the satellite), but on the opposite side of the halo from the satellite ($\approx 180^{\circ}$).

Figure \ref{fig:orientation} shows the angle between the $n=0$ dipole harmonic and the pericentric position at the LMC. As expected, following the passage through the pericenter, the dipole 
appears anti-aligned with the pericentric position. This anti-alignment lasts for $\sim 1-2$ Gyr, 
in line with the length of the dipole response measured from the power (Figure \ref{fig:power}). 
It also corresponds with our expectation that the halo response to the satellite largely dominates the dipole during the LMC's orbital passage (which manifests as the collective response, an overdensity in the outer halo, as discussed in \citealt{Weinberg98, Gomez16, GC21}). 

\begin{figure*}
  \includegraphics[width=\textwidth]{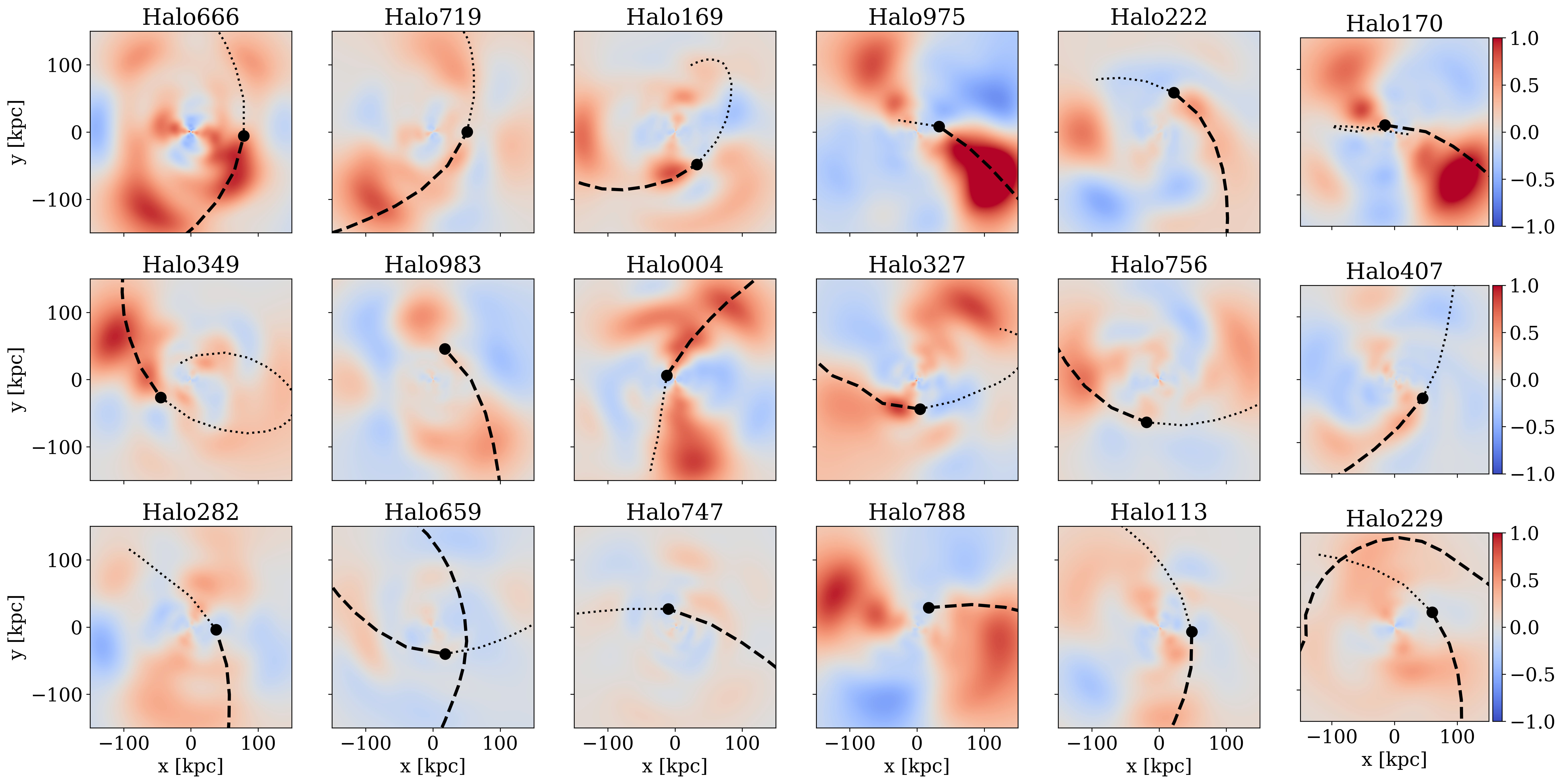}
  \caption{Dynamical friction wakes for the 18 \mwest\ halos. The axes are rotated to show the plane of the LMC analog's orbit, and the density is averaged over 
   $-25$~kpc $< z < 25$~kpc relative to the $l=0$ mode. The panels are sorted by decreasing merger 
   ratio (top--bottom and right--left). The black line indicates the orbit of the LMC analog (dashed \emph{before} pericenter and dotted \emph{after} pericenter), and the black dot 
   shows the current position of the LMC analog (\emph{at} pericenter).} 
  \label{fig:classical_wakes}
\end{figure*}

\subsection{The Classical Dynamical Friction Wake}~\label{sec:classical_wake}

\begin{figure}[ht]
  \includegraphics[width=\columnwidth]{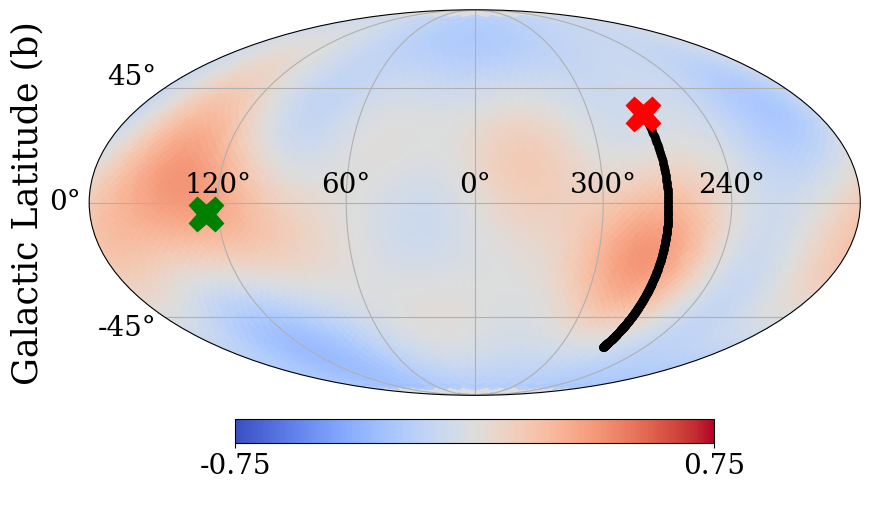}
  
  \caption{Mollweide plot of density contrast at pericenter for Halo407 in the \mwest\ suite, rotated to show the LMC at approximately the present-day position of the LMC. The \emph{red X} marks the position of the LMC, while the \emph{green X} marks the direction of the $l=1$ mode. The black dots show the orbital path of the LMC between infall and pericenter.} 
  \label{fig:mollweide}
\end{figure}

\begin{figure*}
  \includegraphics[width=\textwidth]{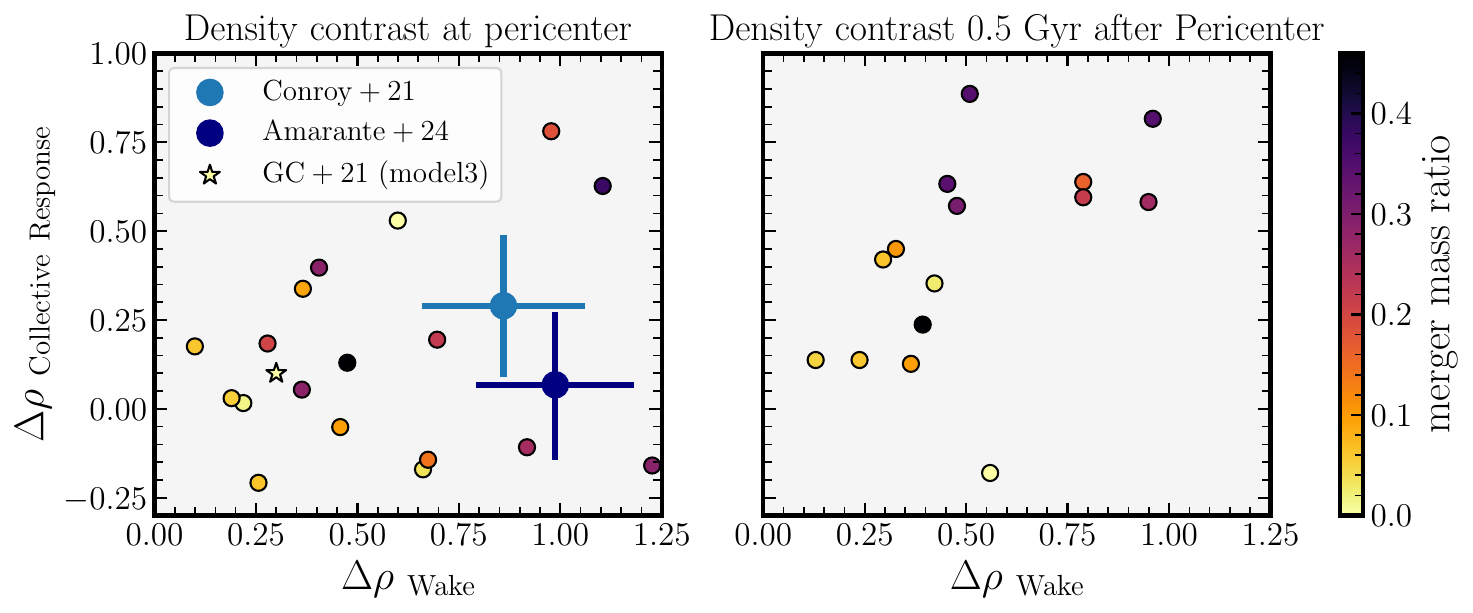}
  
  \caption{DM density contrast in the dynamical friction wake vs. the collective response, measured at pericenter (\emph{left}) and at 0.5 Gyr after 
  pericenter (\emph{right}). For both plots, the density contrast (equation \ref{eq:delta_rho}) is averaged between $60$ kpc $< d < 100$ kpc and measured along 
  the LMC orbital path for the wake, and in the direction of the $l=1$ harmonic for the collective response (Figure \ref{fig:mollweide}). The points are colored
  by $M_{\rm LMC}/M_{\rm MW}$. Also shown in the left panel is the density contrast at pericenter from both the idealized simulations \citep{garavito_19} (colored by the 
  merger ratio) and the observed data \citep{Conroy_2021, Amarante_24}, for the real MW--LMC system, where the LMC is currently roughly at pericenter. At pericenter, there is not a strong correlation (the Spearman correlation coefficient is 0.11) between the overdensity in the wake and the collective response (e.g., cases exist with large overdensity in the wake, but minimal overdensity in the collective response). There is also no correlation with the mass ratio at pericenter. However, the right hand panel indicates that these correlations are stronger 0.5 Gyr after pericenter (the Spearman correlation coefficient is 0.62), when the dipole peaks in strength. The observational results for the present-day MW--LMC system are within the scatter of the simulation results in the left panel.}  
  
  \label{fig:density_contrast}
\end{figure*}

\citet{Chandrasekhar1943} first showed that a point-mass source moving through an infinite medium
would experience a dynamical friction force from the density wake induced by the satellite \citep{Kalnajs70, Mulders83}. This overdensity, which we call here the \textit{dynamical friction wake}, can be seen in Figure~\ref{fig:classical_wakes}, where most of the halos show a density enhancement trailing 
the orbit of the satellite. Similar to Figure~\ref{fig:bfe-decomposition}, the density projections are rotated to match the orbital plane of the satellite. The top left panel corresponds to the larger merger ratios, while the bottom right panel represents the lower mass merger ratios.

In addition to the dynamical friction wake, several overdensities and underdensities 
are also seen in the halo. For example, \textsc{Halos 719, 975, 170, 349, 004, 327, 756, 282, 788, 113}, and \textsc{229} show strong quadrupoles that include a large overdensity opposite to the satellite's location, which is referred to in the literature as the {\it collective response}. 

As further discussed in Section~\ref{sec:power}, there are a number of factors, such as merger 
ratio, halo shape, eccentricity, and pericentric distance, that can modulate the amplitude 
of the halo response. In particular, the density in the dynamical friction wake increases 
with the mass of the satellite, while the collective response depends on the amount of 
displacement induced in the host halo. Here, we compare both the amplitude in the dynamical 
friction wake and in the collective response in all the halos.

To compare the relative overdensity caused by the dynamical friction wake and the 
collective response, we compute the density contrast $\Delta \rho$ defined as: 
\begin{equation}
  \Delta \rho = \rho / \rho_{l=0} - 1,
\end{equation}\label{eq:delta_rho}
where $\rho_{l=0}$ is the density of the monopole. 
We compare the density contrast between 
the halo in the direction of the $l=1$ harmonic (collective response) and along the orbit of the 
satellite (dynamical friction wake). Similar to what is done in observations, we average the density contrast 
(density relative to the monopole) from $60$ kpc $< d < 100$ kpc. We measure the density contrast 
of the wake as the maximum along the orbital path of the LMC and the density contrast of the 
collective response as the maximum value in the direction of the $l=1$ harmonic using a Mollweide 
projection with $N_{\rm side} = 24$ and $1^\circ$ Gaussian smoothing. An example can be seen in 
Figure~\ref{fig:mollweide}, where we have rotated the axis to match the present-day position of the 
LMC in the MW. We measure the collective response at the location of the green X, while the wake is 
measured as the maximum contrast along the black path. 

In Figure~\ref{fig:density_contrast}, we compare the density contrast of the dynamical friction wake
vs. the collective response in our simulations to measurements from the MW \citep{Conroy_2021} 
and idealized simulations \citep{garavito_19}. We show both the values measured at pericenter 
($\sim$ the present-day location of the LMC) and the value at $t-t_{\rm peri} = 0.5$\,Gyr 
(roughly the position where we expect the collective response to reach its maximum value; 
see Figure~\ref{fig:power}). 

At pericenter, our simulations show a wide range of scatter, spanning the values 
measured from both the observed data and the previous simulations. The observational data is not at odds with CDM expectations. However, we do not find a correlation (the Spearman correlation coefficient is
0.11) between the strength of the density contrast in the wake and the density contrast in the collective response. The wide range of densities contrast measured both in the wake and in the collective response 
highlights the non-linear response of the density field of the halo. Hence, constraining the mass of the real LMC from the present-day density contrast is not straightforward, and more sophisticated methods as those presented in \cite{Brooks25b, Brooks25c} are required.

At $t-t_{\rm peri} = 0.5$ Gyr, we see a positive correlation (the Spearman correlation coefficient is
0.62) between the strength of the density contrast in both the dynamical 
friction wake and the collective response. As expected, the density contrast for the collective response is higher, on average, than at pericenter, and shows a 
stronger correlation with the merger ratio. This is in line with recent measurements of a 
present-day density contrast consistent with zero for the collective response \citep{Amarante_24} 
and indicates that the density contrast in the MW caused by the collective response will 
likely grow over the next $\sim 500$\,Myrs, assuming that the present-day LMC is at or near the first 
pericenter.

\section{Discussion}\label{sec:discussion}

\subsection{What sets the power of the dipolar and quadrupolar response at pericenter?} \label{sec:power}

To first order, we expect the power of the response to scale with the tidal force exerted by 
the LMC. A proxy for the tidal force is the scaled tidal index ($\Gamma$), which is calculated as:

\begin{equation}
  \Gamma = \log_{10} \frac{M_{\rm LMC}/d^3}{V_{\rm c,max}^2/GR_{\rm max}^2},
\end{equation}

where $M_{\rm LMC}$ is the mass of the LMC, $d$ is its galactocentric distance, $V_{\rm c,max}$ is the maximum circular velocity of the host, and $R_{\rm max}$ is the corresponding radius. This gives the ratio 
of the tidal forces between the MW and LMC, which is maximized at pericenter. However, this value 
neither accounts for the length of the interaction, which depends on the satellite velocity, 
nor does it capture the full dynamical interaction between the two bodies, which will also depend, e.g., on the orbital distribution in the host. 

We calculate the tidal index as the maximum value using the orbital path of the LMC and the 
instantaneous mass. 
This parameter is relatively noisy due to difficulties in assigning a bound mass at pericenter. These difficulties arise because subhalos receive a strong 
impulsive shock at pericenter \citep{gnedin_1999}, with a median subhalo having nearly twice its 
binding energy rapidly injected into it \citep{vdb_ogiya_2018}. The subhalo remains intact 
because the energy is not evenly distributed across its particles \citep{vdb_ogiya_2018}, 
but these shocks can lead to complex temporary configurations of the subhalo's particles that 
can cause some subhalo finders to incorrectly measure the bound mass. Therefore, we also examine the correlation 
between the power and the pericentric distance ($d_{\rm peri}$) and the infall merger ratio 
($M_{\rm LMC}/M_{\rm MW}$) as well. 

To compute the maximum power in the $l=1$ and $l=2$ harmonics, we first subtract off the background 
power at $t-t_{\rm peri}=-0.5$~Gyr. The power in the $l=1$ and $l=2$ harmonics after this subtraction 
can be seen in Figure~\ref{fig:power} (\emph{bottom}), where the lines are colored by the LMC merger 
ratio. We compute the maximum power as the maximum offset subtracted power between 
$0 < t-t_{\rm peri} < 1.5$~Gyr for the $l=1$ harmonic and between $-0.5 < t-t_{\rm peri} < 0.5$~Gyr 
for the $l=2$ mode. For this analysis, we remove two halos: \textsc{Halo 788}, where the LMC has not yet 
reached pericenter, and \textsc{Halo 983}, which has a massive merger that reaches pericenter $\approx 1$~Gyr 
before the LMC.

\begin{deluxetable}{c S[table-format=1.2] S[table-format=1.2]}
\tablecolumns{3}
\tablewidth{0pt}
\tablecaption{Spearman correlation coefficients\label{tab:spearman}}
\tablehead{
\colhead{Property} & \multicolumn{1}{c}{$\ell=1$ (Dipole)} & \multicolumn{1}{c}{$\ell=2$ (Quadrupole)}
}
\startdata
$b/a$           & \snum{-0.34 $\pm$ 0.24} & \snum{-0.66 $\pm$ 0.19} \\
$c/a$           & \snum{-0.23 $\pm$ 0.26} & \snum{-0.62 $\pm$ 0.19} \\
$d_{\rm peri}$      & \snum{0.08 $\pm $0.28} & \snum{-0.02 $\pm$ 0.31} \\
$M_{\rm LMC}/M_{\rm MW}$ & \snum{0.92 $\pm$ 0.05} & \snum{0.71 $\pm$ 0.10} \\
$T$            & \snum{0.18 $\pm$ 0.27} & \snum{0.36 $\pm$ 0.26} \\
$\Gamma$         & \snum{0.78 $\pm$ 0.11} & \snum{0.62 $\pm$ 0.13} \\
\enddata
\tablecomments{Correlations are measured between normalized gravitational energy at pericenter 
(Figure~\ref{fig:power}) and various halo and orbital properties: the halo axis ratios ($b/a$, $c/a$), 
the pericentric distance ($d_{\rm peri}$), the LMC-to-MW mass ratio ($M_{\rm LMC}/M_{\rm MW}$), 
the triaxiality ($T$), and the scaled tidal index ($\Gamma$). 
Uncertainties are estimated via jackknife resampling.}
\end{deluxetable}

Spearman's correlation coefficient between each of these properties and the maximum power of the 
$l=1$ and $l=2$ harmonics can be seen in Table \ref{tab:spearman}. We see the strongest correlation 
between the power and the merger ratio, followed by the tidal index. We plot the correlation with 
the merger ratio in Figure \ref{fig:strength}. The power in the $l=2$ harmonic appears 
to be moderately correlated with the merger ratio, while the power in the $l=1$ harmonic shows a 
strong correlation. We also plot the normalized power in both the $l=1$ and $l=2$ harmonics from the 
idealized simulations. The maximum power is in agreement with that of the cosmological simulations 
given the merger ratio of the LMC remnant \citep[$M_{\rm LMC}/M_{\rm MW} = 0.12$;][]{garavito_19}

The pericenter shows no correlation with the gravitational energy in either the $l=1$ or $l=2$. However, this may be 
attributable to \mwest\ LMC analogs being constrained to have their pericenters at or around 50~kpc, 
meaning that the suite probes a small range of pericentric distances.

We also examine the correlation between the power in the $l=1$ and $l=2$ and the shape of the 
halo ($b/a$, $c/a$, and the triaxiality (T) as described in Section~\ref{sec:methods}), all 
calculated as the median between $-3\,$ Gyr $< t-t_{\rm peri} < -1\,$ Gyr. We average over this 
timescale to get a proxy for the halo shape prior to the timescales where we expect additional 
deformation induced by the LMC (see Figure \ref{fig:power}). %harmonics and the tidal index ($\Gamma$), halo axis ratios (averaged over $-3 \rm Gyr < t-t_{\rm peri} < -1 \rm Gyr$ prior to infall), triaxiality (T), and pericenter distance. 
Both $b/a$ and $c/a$ are anti-correlated with the strength of the power in the $l=2$ harmonic 
(Table~\ref{tab:spearman}). This is consistent 
with the correlations found in Figure~\ref{fig:correlations} discussed in Section~\ref{sec:correlations} where halos that
are more stretched (lower values of the
axis ratios $c/a$) have higher amplitudes in the 
gravitational energy of the quadrupoles.
This can also be seen in Figure \ref{fig:strength}, where we color the 
points by $c/a$. At fixed $M_{\rm LMC}/M_{\rm MW}$, halos with smaller axis ratios show higher 
maximum power in $l=2$. These results support the results found by \cite{Arora25} that the preexisting triaxiality of the halo can amplify the $l=2$ response to the satellite.

\subsection{Future work: BFE as a framework to characterize and decompose the dynamics of DM halos 
in cosmological simulations}

We have presented a BFE-based characterization of the dynamical response of DM halos 
to the passage of satellites. Representing large-scale features in simulations using BFE provides a natural language for describing simulations and connecting to perturbation theory.
Each term in the BFE is a time series that represents fully or partially the dynamical properties of and processes that 
take place in the halo. For example, in Figure~\ref{fig:correlations} we show that the amplitude of the 
monopole $l=0$ represents the enclosed mass of the halo. BFE therefore offers a powerful framework 
to succinctly characterize and compare the dynamics of halos across simulations. 

In cosmological simulations, unlike idealized $N$-body simulations, many processes drive galaxy evolution simultaneously. This complicates efforts to decompose the response of a halo
to a particular physical mechanism. For example, we showed that the amplitude of the quadrupole harmonic ($l=2$) 
is correlated with the shape of the halo, but can also be excited by satellite galaxies. In this example, it is not clear how to disentangle the halo response from its intrinsic shape. 

In upcoming work, we plan to show how one can analyze BFE with time-series data analysis methods
to decompose the halo response. In \cite{Arora25}, we decompose the halo response to 
filamentary accretion and to the accretion of a massive satellite using multichannel singular spectral analysis
on the time series of the $l=2$ modes. In Varela et. al., in prep, we decompose the torques experienced
by a galactic disk into the passage of a satellite and to the DM halo response of the host. 

\section{Conclusions}\label{sec:conclusions}

In this work --- the second paper of a three-part series investigating the dynamics of Milky Way (MW)-like halos in cosmological simulations --- we examine the halo response of MW-like galaxies to LMC-like satellites in a cosmological context. We use the \mwest\ suite of zoom-in cosmological simulations, which includes 18 analogs of the MW–-LMC system~\citep{buch2024milky}. These simulations are well-suited for this study, as they broadly reproduce the expected infall mass of the LMC and its orbital properties, particularly a pericenter of 30--70 kpc within the past Gyr. 
As a comparison sample, we use a subset of the \textsc{Symphony MilkyWay} simulations that did not experience any major mergers or LMC-like mergers within the past 5 Gyr.

We employ Basis Function Expansions (BFE) as a framework to analyze the halo response, providing a natural language for interpreting the structure and evolution of halos \citep{EXP}. BFE enables consistent quantification of differences in halo responses across the simulation suite. Here, we use BFE to identify and characterize the evolution of the host halo response caused by the recent pericentric passage of an LMC-like satellite in 18 \mwest\ hosts, 8 \symphony halos that do not present mergers, and compare them with the 8 idealized MW--LMC $N$-body simulations presented in \cite{GC21}. We decompose the halo response in terms of the MW--LMC mass ratio, halo density center displacement, and MW halo shape, using the amplitude and evolution of the dipole and quadrupole terms of the BFE. Our main findings are as follows:

\begin{enumerate} 
  \item \textbf{The halo response in all 18 \mwest\ halos is dominated by
  the dipole and the quadrupole (Figure~\ref{fig:all-mean-modes})}. The 
  dipole captures the displacement induced by the LMC-like satellite, 
  while the quadrupole reflects both the host's initial halo shape and
  the formation of the dynamical friction wake tracing the satellite's orbit.

  \item \textbf{Dynamical friction wakes are present in all halos with 
  LMC-like satellites}. We identify dynamical friction wakes in all 18
  \textsc{MWest} halos. Their amplitudes, which trail the LMC's orbit,
  peak right before the first pericentric passage of the satellite, and
  the peak density scales with the satellite mass (Figure~\ref{fig:density_contrast}).
  However, the wake morphologies are more complex than in idealized halos,
  owing to the underlying halo triaxiality, which affects the response
  (Figure~\ref{fig:strength}, right panel). 

  \item \textbf{Evolution, amplitude, and physical meaning of the dipole ($l=1$):}
  The dipole harmonic captures the response of the host to the LMC-like satellite 
  (middle panel in Figure~\ref{fig:correlations}). The amplitude 
  of the dipole is controlled by the MW--LMC mass ratio (left panel in Figure~\ref{fig:strength}). 
  The peak dipole amplitude occurs $\approx 0.2-0.7$~Gyr after the satellite's first pericenter
  (Figure~\ref{fig:power}), suggesting that the dipole is still increasing in strength in the 
  real MW--LMC system and consistent with idealized simulations. However, the amplitude of the dipole 
  in idealized simulations is $\approx$20 times smaller, most likely due to the absence of environmental perturbations in the idealized simulations. The dipole direction (Equation~\ref{eq:dipole_angles}) lies consistently opposite to the LMC-analog pericenter on the sky and is aligned with the plane of the orbit 
  (Figure~\ref{fig:orientation}). In contrast, dipoles are completely absent 
  in the 8 \symphony halos (upper left panel in Figure~\ref{fig:power}).
  
  \item \textbf{The quadrupole ($l=2$) mainly characterizes the host halo triaxiality, with an additional contribution from the dynamical friction wake}.
  All host halos show a persistent quadrupole over long periods, including the \symphony\ halos, which lack a recently accreted massive satellite. We found that the quadrupole primarily reflects the host halo's triaxiality (Figure~\ref{fig:correlations}).
  However, we observe a peak in the quadrupole amplitude near the LMC analog's pericenter (Figure~\ref{fig:power}), generated by the dynamical friction wake. Although subdominant to the initial amplitude of the quadrupole, this peak is at least twenty times as large as in the idealized simulations
  (which are absent of triaxiality and environmental perturbations). The amplitude of the quadrupole at pericenter also shows a secondary correlation with the axis ratios of the host, indicating that preexisting triaxiality may enhance the halo response to the satellite.  

  \item \textbf{No strong correlation between the density contrast of the dynamical friction wake and the collective response at satellite pericenter}. 
  The collective response refers to the overdensity in the outer halo caused by the displacement of the host by the satellite.
  About 0.5 Gyr after the satellite's pericenter, when the dipole is maximized, the amplitude of both the 
  collective response and the dynamical friction wake increase with the MW--LMC mass ratio. 
  However, the real LMC is currently at a pericentric approach to the MW. At pericenter, we find no correlation between the two density contrasts or trend with MW--LMC mass ratio. The host halo shape prior to the satellite infall (characterized by the quadrupole) induces scatter in the density contrast. This scatter obscures any trend with MW--LMC mass ratio. 
  %is too large to constrain the actual merger mass ratio. 
  This scatter was not observed in idealized $N$-body simulations, which typically assume a spherical MW halo. Interestingly, at pericenter, the measured density contrast in the dynamical friction wake and the collective response span the range of values previously measured in both idealized simulations and observations of the stellar halo (Figure~\ref{fig:density_contrast}), indicating consistency with CDM expectations.
  
  \item \textbf{Disentangling the present-day MW halo response to recover the pre-infall structure:}
  In CDM theory, the MW's initial halo shape prior to the LMC's infall encodes information about its assembly history. However, the strong halo response to the LMC's recent infall complicates recovery of this initial distribution. We find that the dipole and quadrupole contain complementary information: the dipole amplitude is dominated by the MW's halo density displacement, set by the MW--LMC mass ratio,
  while the quadrupole amplitude reflects the pre-infall halo shape,
  with additional contribution from the dynamical friction wake.

\end{enumerate}

Looking ahead, our results highlight the importance of jointly measuring dipole and quadrupole signatures in the MW halo to disentangle the effects of the LMC from the Galaxy’s pre-infall structure. Upcoming wide-field stellar surveys, including DESI, Gaia, LSST, and Roman will map the outer halo with unprecedented precision, making it possible to detect and characterize these distinct distortions observationally. By connecting the observed dipole and quadrupole imprints to simulations, we can begin to recover the MW’s initial halo shape and assembly history, and more broadly, establish satellite--host interactions as a key probe of dark matter and galaxy formation physics.

\section*{Acknowledgments}
This work used data from the \mwest\ and \symphony\ suite of simulations, hosted at \href{https://web.stanford.edu/group/gfc/gfcsims/}{web.stanford.edu/group/gfc/gfcsims}. This work was supported by the Kavli Institute for Particle Astrophysics and Cosmology at Stanford University, SLAC National Accelerator Laboratory, and the U.S. Department of Energy under contract number DE-AC02-76SF00515 to SLAC National Accelerator Laboratory. NGC and KJD acknowledge support provided by the Heising-Simons Foundation grant \# 2022-3927. KVJ's contributions were supported by a grant 1018465 from the Simons Foundation. A.A. acknowledges support from the DiRAC Institute in the Department of Astronomy and the eScience Institute, both at the University of Washington. CL acknowledges funding from the European Research Council (ERC) under the European Union’s Horizon 2020 research and innovation programme (grant agreement No. 852839) and funding from the Agence Nationale de la Recherche (ANR project ANR-24-CPJ1-0160-01). 
GB's contributions were supported by NSF CAREER award AST-1941096 and NASA ATP award 80NSSC24K1225.
SVL acknowledges financial support from ANID/Fondo ALMA 2024/31240070. FAG acknowledges support from the ANID BASAL project FB210003, from the ANID FONDECYT Regular grants 1251493, and from the HORIZON-MSCA-2021-SE-01 Research and Innovation Programme under the Marie Sklodowska-Curie grant agreement number 101086388.
EDF and NGC are grateful to Adrian Price-Whelan, the EXP collaboration, and the Nearby Universe group at the Center for Computational Astrophysics of the Flatiron Institute for valuable discussions and comments that helped shape this paper.
KJD and NGC acknowledge that the University of Arizona is home to the O’odham and the Yaqui.
NGC acknowledges that the University of Maryland is home to the Piscataway. 
We respect and honor the ancestral caretakers of the lands upon which this work was performed, from time immemorial until now, and into the future.

\begin{contribution}
EDG performed all the analyses and wrote the first draft of the paper.
NGC carried out the analysis presented in Section 3.1.2, contributed to the writing, and submitted the manuscript.
AA wrote code to carry out the analysis, edited the manuscript, and provided guidance.
RW and GB edited the manuscript and provided guidance on the analysis.
PM ran symfind and provided the particle data used in the analysis.
MS, MW, and KVJ provided guidance on the analysis of the paper.
DB and EN ran the simulations used in this work.
All authors provided comments on the manuscript at various stages. 
\end{contribution}

\software{
This work made use of the following software packages: \texttt{python} \citep{python}, \texttt{Jupyter} \citep{kluyver2016jupyter}, \texttt{numpy} \citep{numpy}, \texttt{scipy} \citep{scipy},  \texttt{matplotlib} \citep{matplotlib}, \texttt{EXP} \citep{EXP}, \texttt{Symfind} \citep{mansfield2023symfind}, \texttt{pynbody} \citep{Pontzen2013}, \texttt{pytreegrav} \citep{Grudic2021}, \texttt{astropy} \citep{astropy}. 
This research has made use of the Astrophysics Data System, funded by NASA under Cooperative Agreement 80NSSC21M00561.}

\appendix

\section{Constructing a BFE for cosmological halos}\label{sec:appendix}

In this section, we outline the procedure we follow to 
build the BFE for all the \mwest\ halos. As an example, we demonstrate
the process with \textsc{Halo}~349 from the \mwest\ suite. The density field decomposition for this halo was shown in Figure~\ref{fig:bfe-decomposition}. We use as input data the positions and masses of the particles at every snapshot of each halo; at this stage, all the substructure of the halo was removed. The halos are centered on the cusp of the halo and rotated in such a way that the orbital plane of the satellite is in the $x-y$ plane as described in Section~\ref{sec:orientation}. All the analyses presented here were performed using \textsc{pyEXP} v7.7.99.

\begin{figure}[ht]
  \centering
  \includegraphics[width=1.0\linewidth]{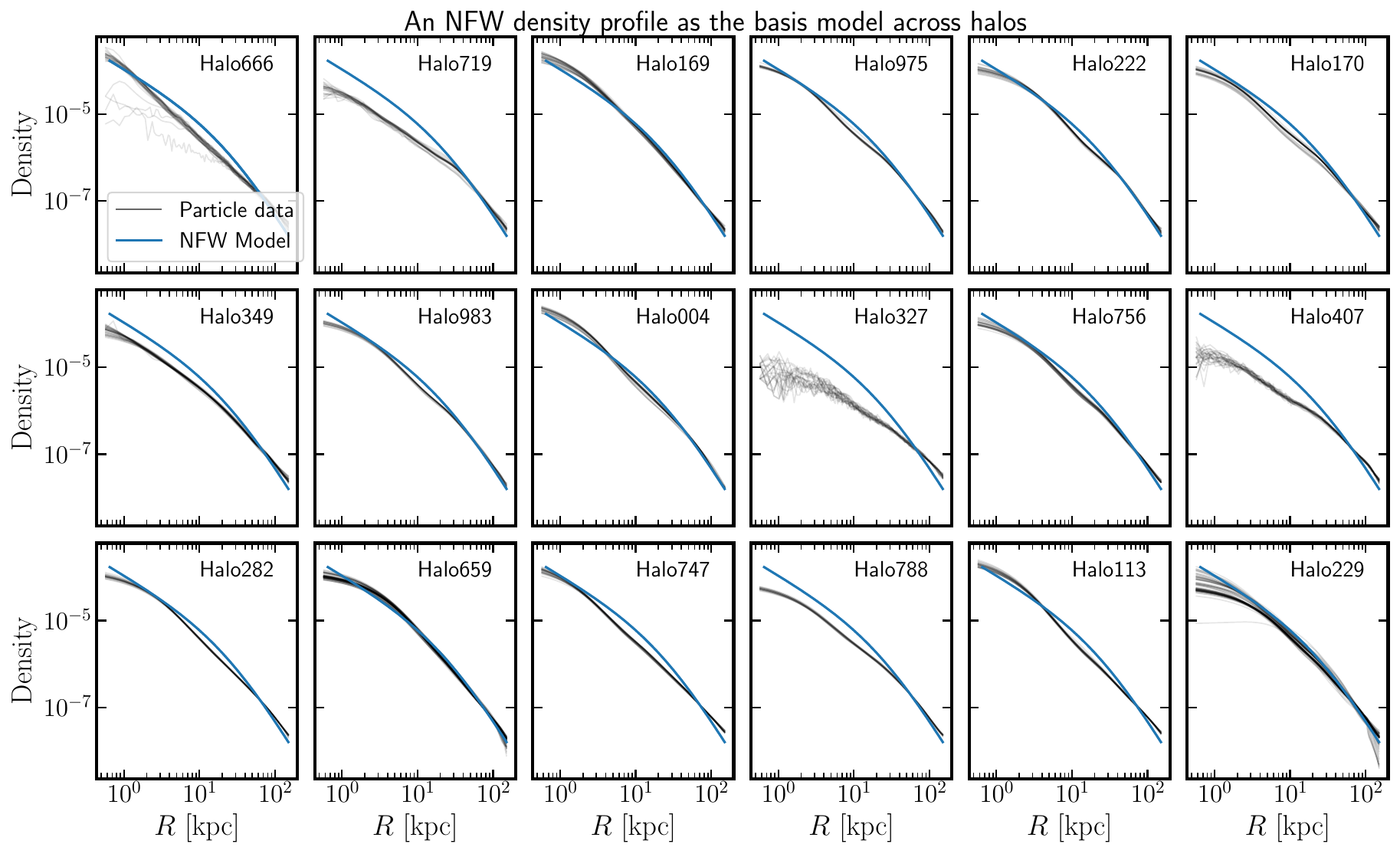}
  \caption{Density profiles of all the halos of the \mwest\ simulations. Black lines represent the density profile at each snapshot in the simulation from $z=1-0$. The blue line shows the best fit NFW profile for all 18 halos at every time. This is the profile that is used to build the basis.}\label{fig:nfw_basis}

\end{figure}

\begin{figure}
  \centering
  \includegraphics[width=0.8\linewidth]{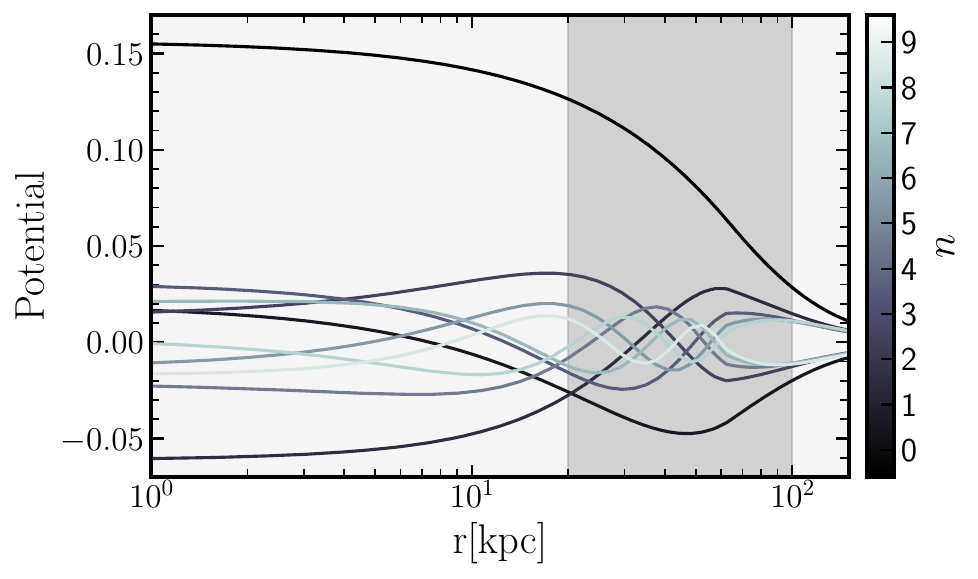}
  \caption{Radial profile of the basis as a function of $n$. The $y$-axis represents the gravitational potential profile computed from the basis assuming all the coefficients are unity. The $n=0$ harmonic represents the zeroth-order NFW basis. In the higher $n$-orders, a higher nodal structure can be seen, in particular at the distance highlighted by the shaded grey region. The nodes of the basis are not evenly distributed in radius; most of the nodes are between 10--100~kpc. It is in these regions where we expect to resolve most of the structure of the halos.}
  \label{fig:basis}
\end{figure}

\begin{figure}
  \centering
  \includegraphics[width=0.5\linewidth]{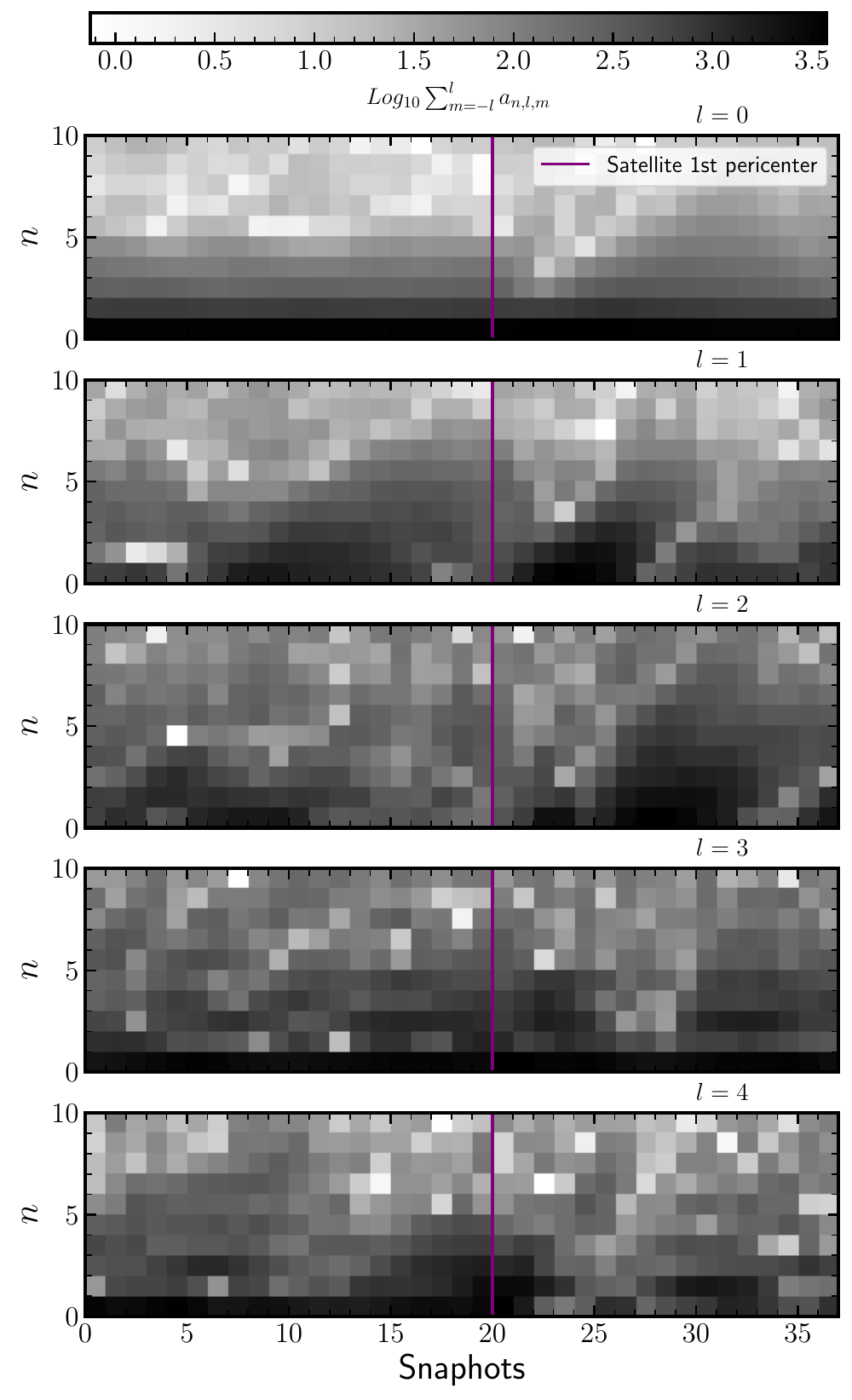}
  \caption{Visualization of the amplitude in the coefficients of \textsc{Halo}~349. Each subplot corresponds to a different $l$ harmonic mode, and each pixel is the sum over all the available $m$ harmonics for a given $n$ and $l$ value. The vertical purple line corresponds to the satellite's pericentric passage. The satellite passage induces perturbations in the structure of the halo that can be seen as amplitude variations in the coefficients before and after the pericentric passage.}
  \label{fig:coefs}
\end{figure}

\begin{enumerate}
  \item \textbf{Make the Basis Model: Choosing the Zeroth-Order Basis}.\\ Perhaps one of the most critical steps is to choose the zeroth-order basis. In \textsc{EXP}, one can compute an empirical basis or one can 
  use an analytic density profile that fits the halo's density as the zeroth-order basis function.  Subsequent basis functions in the series are orthogonal to the previous functions and add additional nodes that represent spatial variations of decreasing linear scale. In our case, since we aim to compare 18 halos, it is ideal to adopt a common basis across all halos. The best fit for our set of halos is an NFW profile with a scale length of 25~kpc shown with the solid blue line in Figure~\ref{fig:nfw_basis}. This profile provides a good fit for most of the halos; however, in a couple of cases ( \textsc{Halo} 719, \textsc{Halo} 327, and \textsc{Halo} 407), the density profiles within $\approx 20$~kpc are not a good fit. Nonetheless, the higher-order harmonics in the expansion will compensate for the poor fit. In particular, for \textsc{Halo}~327
  we see large dipoles in Figure~\ref{fig:classical_wakes}. 
  The right-hand panel of Figure~\ref{fig:dens_residuals} demonstrates that the higher-order harmonics compensate for a non-ideal fit to yield an accurate density profile. The best-fit model is tabulated in a text file with four columns corresponding to the radius, density, mass, and potential.
  
  \item \textbf{Building the NFW basis:}\\
  Once the model is defined, one can compute the bi-orthogonal basis for the 
  targeted model. In \textsc{pyEXP}, this is done by 
  specifying the model parameters and expansion length in 
  a \textsc{YAML} file, which is then passed as an argument to the \verb|pyEXP.basis.Basis.factory| function. For our model,  we use the following YAML config file:
  
\begin{verbatim}  
  printid : sphereSL
  parameters :
   numr   : 1000
   rmin   : 0.99999
   rmax   : 150
   Lmax   : 5
   nmax   : 10
   rmapping : 0.0166666
   modelname : mwest_nfw.txt
   cachename : mwest_nfw.cache
  \end{verbatim}
  Figure~\ref{fig:basis} shows the shape of the basis as a function of radius and of the radial ($n$) order of the expansion. For $n=0$, the potential is the NFW. The number of functional nodes increases with $n$. However, note that the nodes are not even distributed as a function of radius. Most of the nodes are in the vicinity of the characteristic NFW radius $a$.
  
  \item \textbf{Computing the coefficients:}\\  Lastly, the coefficients are computed using the positions and masses of the particles. This is achieved with the 
  \verb|pyEXP.coefs.Coefs.makecoefs| and the \verb|basis.createFromArray| member functions. It is critical that the 
  halo particles are centered on the desired center of the expansion to avoid artifacts in the higher-order harmonics. The coefficients are then computed at every snapshot in the halo's evolution. The coefficients of \textsc{Halo} 349 are shown in Figure~\ref{fig:coefs}. Visualizing the coefficient's time series provides insight into how halos evolve. For example, after the pericentric passage of the satellite, several coefficients are enhanced (darker regions). Each coefficient also has information about the direction and amplitude of the perturbation.

\end{enumerate}

\setlength{\bibsep}{0pt} % Reduces space between entries
\renewcommand{\baselinestretch}{1.0} % Remove paragraph spacing
\bibliography{references}{}
\bibliographystyle{aasjournalv7}

%\section{Milky Way-est Suite}\label{sec:mwest_suite}

%% the collaboration and author truncation commands are used. It has to


\end{document}